\documentclass[12pt]{article}

\usepackage{amsbsy}
\usepackage{amsfonts}
\usepackage{amssymb}

\usepackage{bm}
\usepackage{enumerate}
\usepackage{graphics}
\usepackage{epsfig}
\usepackage{color}
\usepackage{nicefrac}
\usepackage{mathrsfs}
\usepackage{bbm}
\usepackage{amsmath}
\usepackage{amsthm}
\usepackage{graphics}
\usepackage{epsfig}

\input epsf.sty

\newlength{\TZ}
\newcommand{\BEQ}{\begin{equation}}     % Gleichungen Anfang ..
\newcommand{\BEA}{\begin{eqnarray}}
\newcommand{\BD}{\begin{displaymath}}
\newcommand{\EEQ}{\end{equation}}       % .. und Ende
\newcommand{\EEA}{\end{eqnarray}}
\newcommand{\ED}{\end{displaymath}}
\newcommand{\eps}{\varepsilon}          % epsilon
\newcommand{\vph}{\varphi}              % rundes phi
\newcommand{\D}{{\rm d}}                % gerades d fuer Ableitungen
\newcommand{\II}{{\rm i}}               % gerades i fuer komplexe Einheit
\newcommand{\demi}{\frac{1}{2}}         % Bruch 1/2
\newcommand{\wit}[1]{\widetilde{#1}}    % weite Schlange
\renewcommand{\vec}[1]{\boldsymbol{#1}} % Vektoren fettgedruckt

\newcommand{\appsektion}[1]{\setcounter{equation}{0}\setcounter{subsection}{0}
\section*{Appendix. #1}
\renewcommand{\theequation}{A.\arabic{equation}}
              \renewcommand{\thesection}{A} }
\catcode`\@=11
\def\numberbysection{\@addtoreset{equation}{section}
        \def\theequation{\thesection.\arabic{equation}}}
\numberbysection

\begin{document}

\begin{titlepage}

\vskip 1.5 cm
\begin{center}
{\LARGE \bf Meta-conformal algebras in $d$ spatial dimensions}
\end{center}

\vskip 2.0 cm
\centerline{{\bf Malte Henkel}$^{a,b,c}$ and {\bf Stoimen Stoimenov}$^d$}
\vskip 0.5 cm
\centerline{$^a$ Rechnergest\"utzte Physik der Werkstoffe, Institut f\"ur Baustoffe (IfB), ETH Z\"urich,}
\centerline{Stefano-Franscini-Platz 3, 	CH - 8093 Z\"urich, Switzerland}
\vspace{0.5cm}
\centerline{$^b$ Groupe de Physique Statistique, D\'epartement de Physique de la Mati\`ere et des Mat\'eriaux,}
\centerline{Institut Jean Lamour (CNRS UMR 7198), Universit\'e de Lorraine Nancy,}
\centerline{B.P. 70239, F -- 54506 Vand{\oe}uvre l\`es Nancy Cedex, France\footnote{Permanent address,
after 1$^{\rm st}$ of January 2018:
Laboratoire de Physique et Chimie Th\'eoriques (CNRS UMR), Universit\'e de Lorraine Nancy, B.P. 70239, F - 54506 Vand{\oe}uvre-l\`es Nancy Cedex, France}}
\vspace{0.5cm}
\centerline{$^c$Centro de F\'{i}sica Te\'{o}rica e Computacional, Universidade de Lisboa, P--1749-016 Lisboa, Portugal}
\vspace{0.5cm}
\centerline{$^d$ Institute of Nuclear Research and Nuclear Energy, Bulgarian Academy of Sciences,}
\centerline{72 Tsarigradsko chaussee, Blvd., BG -- 1784 Sofia, Bulgaria}

\begin{abstract}
Meta-conformal transformations are constructed as dynamical symmetries of the linear transport equation in $d$ spatial dimensions.
In  one and two dimensions, the associated Lie algebras are infinite-dimensional and isomorphic to the direct sum of either two
or three Virasoro algebras. Co-variant two-point correlators are derived and possible physical applications are discussed.
\end{abstract}
\end{titlepage}

\setcounter{footnote}{0}

%%%%%%%%%%%%%%%%%%%%%%%%%%%%%%%%%%%%%%%%%%%%%%%%%%%%%%%%%%%%%%%%%%%%%%%%%%%%%%%%%%%%%%%%%%%%%%%%%%%%%%%%%%%%%%%%%%%%%%%%%%%%%%%%%
\section{Introduction}
%%%%%%%%%%%%%%%%%%%%%%%%%%%%%%%%%%%%%%%%%%%%%%%%%%%%%%%%%%%%%%%%%%%%%%%%%%%%%%%%%%%%%%%%%%%%%%%%%%%%%%%%%%%%%%%%%%%%%%%%%%%%%%%%%

Conformal invariance has found many brilliant applications, for example to string theory and
high-energy physics \cite{Polchinski01}, or to two-dimensional phase transitions \cite{Belavin84,Francesco97,Henkel99,Rychkov17}
or the quantum Hall effect \cite{Cappelli93,Hansson17}.
These applications are based on a geometric definition of conformal transformations, considered as local coordinate
transformations $\vec{r}\mapsto \vec{r}'=\vec{f}(\vec{r})$, of spatial coordinates $\vec{r}\in\mathbb{R}^2$
such that angles are kept unchanged.\footnote{See \cite{Rychkov17} and refs. therein for the considerable recent interest into the
case $\vec{r}\in\mathbb{R}^d$ with $d>2$.} The Lie algebra of these transformations is naturally called the `{\it conformal Lie algebra}'.

In view of these successes, it appears natural to ask if at least some ideas of conformal invariance can be brought to bear on dynamical problems.
Indeed, the first attempt we are aware of concerns the critical dynamics of a two-dimensional statistical system \cite{Cardy85}.
In general, the global scaling of time and space coordinates is distinguished by the {\em dynamical exponent $z$}, according to
$t\mapsto t'=b^z t$ and $\vec{r}\mapsto \vec{r}'=b \vec{r}$. In general, $z$ has a non-trivial value \cite{Taeu14}.
Starting from the well-established
conformal invariance in the spatial coordinates, with a generic space-dependent re-scaling factor $b=b(\vec{r})$, for generic $z$,
generalised conformal transformations are derived for both two-time correlators and two-time response functions.\footnote{{\it At} equilibrium, these
are related by the fluctuation-dissipation theorem.}  However, it seems that the absence of supporting physical examples led to the abandon of
this idea. We shall present here the first example where this idea can be implemented, and the associated Lie algebra generators be explicitly
constructed, at least for a dynamical exponent $z=1$.

Concerning symmetry approaches to dynamics in the presence of a dynamic scaling behaviour, it turned out to be more fruitful to rather consider
conformal transformations in {\em time}, e.g. $t\mapsto \frac{\alpha t+\beta}{\gamma t+\delta}$ with $\alpha\delta-\beta\gamma=1$, and then to choose
the spatial transformation such as to obtain a closed Lie group. The free diffusion equation has such a dynamical symmetry, by now usually called the
{\em Schr\"odinger group} \cite{Niederer72}, with dynamical exponent $z=2$. In any space dimension, this non-semi-simple Lie group has an
infinite-dimensional extension, the {\em Schr\"odinger-Virasoro group}, although one usually analyses its Lie algebra \cite{Henkel94,Unterberger12}.
Analogous ideas can be brought forward for generic values of $z$ \cite{Henkel02},
lead to explicit prediction for the scaling form of the two-time response function and have been
tested in a large variety of non-equilibrium systems which undergo dynamical scaling,
see \cite{Henkel10} for a review and \cite{Henkel17c} for a tutorial introduction.

In this work, we shall be interested in systems which undergo dynamical scaling with a dynamical exponent $z=1$. Trivial examples would be given
by conformally invariant critical systems at equilibrium, where one of the spatial direction would be relabeled as `time'. For the sake of
a clear distinction, we refer to the `standard' conformal transformations, which keep angles unchanged and mentioned so far,
as {\it ortho-conformal transformations}. Here, we shall be interested in Lie algebras of time-space transformations, which as Lie algebras
are still isomorphic to the ortho-conformal Lie algebra or at least contain it as a sub-algebra, but which no need to be angle-preserving.
Such transformations will be called {\it meta-conformal transformations} \cite{Henkel17a,Henkel17b}.
To make this idea explicit, consider the infinitesimal generators, acting on a two-dimensional time-space with
points $(t,r)\in\mathbb{R}^2$ \cite{Henkel02}
\BEA
X_n   &=& -t^{n+1}\partial_t-\mu^{-1}[(t+\mu r)^{n+1}-t^{n+1}]\partial_r
          - (n+1)\frac{\gamma}{\mu}[(t+\mu r)^{n}-t^{n}] -(n+1)xt^n\nonumber\\
Y_{n} &=& -(t+\mu r)^{n+1}\partial_r- (n+1)\gamma (t+\mu r)^{n}
\label{infinivarconf}
\EEA
where $x,\gamma$ are constants and $\mu^{-1}$ is a constant universal velocity (`speed of sound or speed of light').
The global dilatations are generated by $X_0$ and it is easy to see that indeed $z=1$.
Clearly, these infinitesimal transformations are not angle-preserving in the time-space $(t,r)\in\mathbb{R}^2$,
but their Lie algebra $\langle X_n, Y_n\rangle_{n\in \mathbb{Z}}$ obeys
\BEQ
[X_n,X_{m}] = (n-m)X_{n+m},\quad  [X_n,Y_{m}] = (n-m)Y_{n+m},\quad [Y_n,Y_{m}] = \mu (n-m)Y_{n+m}
\label{commutators}
\EEQ
The isomorphism of (\ref{commutators}) with the ortho-conformal Lie algebra can be seen by writing
$X_n=\ell_n +\bar{\ell}_n$ and $Y_n=\mu^{-1}\bar{\ell}_n$.
Then the generators $\langle \ell_n, \bar{\ell}_n\rangle_{n\in \mathbb{Z}}$ satisfy
$[\ell_{n},\ell_m]=(n-m)\ell_{n+m}, [\bar{\ell}_n,\bar{\ell}_m]=(n-m)\bar{\ell}_{n+m}, [\ell_n,\bar{\ell}_m]=0$.
Provided $\mu\ne 0$, the Lie algebra (\ref{commutators})
is isomorphic to a pair of Virasoro algebras
$\mathfrak{vect}(S^1)\oplus\mathfrak{vect}(S^1)$ with a vanishing central charge \cite{Henkel02,Stoimenov15}.

The meta-conformal generators (\ref{infinivarconf}) are dynamical symmetries of the equation of motion
\BEQ \label{ineq1}
{\hat S}\phi(t,r)=(-\mu\partial_t+\partial_r)\phi(t,r)=0.
\EEQ
Indeed, since (with $n\in\mathbb{Z}$)\index{conformal invariance}
\BEQ \label{dynsym}
{} [{\hat S},X_n] = -(n+1) t^n {\hat S} +n(n+1)\mu \left( x -\frac{\gamma}{\mu}\right)t^{n-1} \;\; , \;\;
{} [{\hat S},Y_{n}] = 0
\EEQ
a solution $\phi$ of  with scaling dimension $x_{\phi}=x=\gamma/\mu$ is mapped onto another solution of (\ref{ineq1}).
Hence the space of solutions of the equation (\ref{ineq1}) is meta-conformally invariant. This is the analogue of the
familiar ortho-conformal invariance of the $2D$ Laplace equation.

Several different types of physical systems with dynamical exponent $z=1$ are known. First, the dynamical symmetries of the
Jeans-Vlassov equation \cite{Jeans1915,Vlasov38,Henon82,Mo10,Vilani14,Campa09,Campa14,Elskens14,Pegoraro15}
in one space dimension are given by a representation of (\ref{commutators}),
distinct from (\ref{infinivarconf}) \cite{Stoimenov15}. Second, the `non-relativistic limit' $\mu\to 0$ in the above generators
yields a Lie-algebra contraction of (\ref{commutators}) whose result is called `conformal galilean algebra' $\mbox{\sc cga}(d)$
\cite{Havas78} or `{\sc bms}-algebra' $\mathfrak{bms}_{2+d}$\cite{Bondi62},
since the contracted generators are immediately generalised to $d\geq 1$ spatial dimensions, and rotations by arbitrary time-dependent angles
appear for $d\geq 2$. Remarkably, the Lie algebra $\mbox{\sc cga}(d)$
is not isomorphic to the Schr\"odinger (or Schr\"odinger-Virasoro) Lie algebra in $d$ dimensions \cite{Henkel03a,Duval09}.
Applications arise in hydrodynamics \cite{Zhang10} or in gravity, e.g. \cite{Barnich07,Bagchi09,Bagchi10,Martelli10,Barnich13,Bagchi13b},
and the bootstrap approach has been tried \cite{Martelli10,Bagchi17}. Third, the non-equilibrium dynamics of quantum quenches generically
has $z=1$, related to ballistic spreading of signals, see \cite{Calabrese07,Calabrese16,Dutta15}
and this apparently holds both for quenches in the vicinity of the quantum critical point
\cite{Delfino17} as well as for deep quenches into the two-phase coexistence region \cite{Wald17}. The available examples suggest that the
value $z=1$ should be robust with respect to the change from closed to open quantum systems. Forth, effective equations of motion of the form
(\ref{ineq1}) arise in recent studies of the generalised hydrodynamics required for the description of strongly interacting non-equilibrium
quantum systms \cite{Bertini16,Castro16,Doyon17,Caux17,Piroli17,Dubail16}.

%%++++++++++++++++++++++++++++++++++++++++++++++++++++++++++++++++++++++++++++++++++++++++++++++++++++++++++++++++++++++++++%%
\begin{table}
\begin{center}\begin{tabular}{|l|lll|l|} \hline
group                    & \multicolumn{3}{l|}{coordinate changes}                                                  & co-variance \\ \hline
ortho-conformal $(1+1)D$ & $z'=f(z)$      & $\bar{z}'=\bar{z}$                        &                             & correlator  \\
                         & $z'=z$         & $\bar{z}'=\bar{f}(\bar{z})$               &                             & \\ \hline
Schr\"odinger-Virasoro   & $t'=\beta(t)$  & \multicolumn{2}{l|}{$\vec{r}'=\left(\D\beta(t)/\D t\right)^{1/2} \vec{r}$} & response \\
                         & $t'=t$         & $\vec{r}'=\vec{r}+\vec{a}(t)$             &                             & \\ \hline
conformal galilean       & $t'=\beta(t)$  & \multicolumn{2}{l|}{$\vec{r}'=\left(\D\beta(t)/\D t\right) \vec{r}$}    & correlator \\
                         & $t'=t$         & $\vec{r}'=\vec{r}+\vec{a}(t)$             &                             & \\ \hline
meta-conformal $1D$      & $t'=f(t)$      & $\rho'=f(\rho)$                           &                             & correlator \\
                         & $t'=t$         & $\rho'=a(\rho)$                           &                             & \\ \hline
meta-conformal $2D$      & $t'=t$         & $z'=f(z)$                                 & $\bar{z}'=\bar{z}$          & \\
                         & $t'=t$         & $z'=z$                                    & $\bar{z}'=\bar{f}(\bar{z})$ & \\
                         & $t'=\theta(t)$ & $w'=w$                                    & $\bar{w}'=\bar{w}$          & \\ \hline
\end{tabular}\end{center}
\caption[tab1]{Several examples of infinite-dimensional groups of time-space transformations, with the defining coordinate changes.
Herein, $f,\bar{f},\theta$ are arbitrary functions, and $\vec{a}$ an arbitrary vector-valued function, of their argument.
In addition, $z,\bar{z}$ are (complex) light-cone coordinates, $\rho=t+\mu r$ and $w=t+\beta z$, $\bar{w}=t+\beta \bar{z}$.
The physical interpretation of the co-variant $n$-point functions as either correlators or responses is based on the extension of
the Cartan sub-algebra \cite{Henkel14a,Henkel15,Henkel16}. \label{tab1}}
\end{table}
%%++++++++++++++++++++++++++++++++++++++++++++++++++++++++++++++++++++++++++++++++++++++++++++++++++++++++++++++++++++++++++%%

In table~\ref{tab1} we collect the coordinate transformations of these examples of infinite-dimensional groups of time-space transformations.
While the physical interpretation of the co-variant $n$-point functions of ortho-conformal invariance as correlators is long-established, for the
other groups a systematic extension of the Cartan sub-algebra must considered \cite{Henkel03a,Henkel15,Henkel16}. This yields either causality
conditions appropriate for a response function or symmetry conditions as expected for a correlator (this construction has not yet been carried out
for the meta-conformal transformations in two dimensions).\footnote{Furthermore, without these extensions, i.e.
the na\"{\i}ve co-variant two-point functions become
$\langle\phi(t,r)\phi(0,0)\rangle_{\mbox{\scriptsize \sc cga}} = \Phi^{(0)} t^{-x} \exp\left[ - \xi r/t\right]$  for the conformal galilean case and
$\langle\phi(t,r)\phi(0,0)\rangle_{\mbox{\scriptsize\rm meta}} = \Phi^{(0)} t^{-x} \left[ 1 + \mu r/t\right]^{-\xi/\mu}$
for the $1D$ metaconformal case. These show non-physical singularities.
Extending the Cartan algebra modifies the results to the physically acceptable forms
$\langle\phi(t,r)\phi(0,0)\rangle_{\mbox{\scriptsize \sc cga}} = \Phi^{(0)} t^{-x} \exp\left[ - \xi |r/t| \right]$ and
$\langle\phi(t,r)\phi(0,0)\rangle_{\mbox{\scriptsize\rm meta}} = \Phi^{(0)} t^{-x} \left[ 1 + \mu |r/t|\right]^{-\xi/\mu}$ \cite{Henkel16}.
Bootstrap approaches should reproduce these singularity-free forms.} The additional time-dependent rotations of the Schr\"odinger and
conformal galilean groups for $d\geq 2$ dimensions are not explicitly listed. Only the ortho-conformal transformation include rotations between the
`time' and `space' coordinates.

Here, we wish to investigate the existence of meta-conformal dynamical symmetries in more than one spatial dimension.
We shall therefore look for dynamical symmetries of higher-dimensional analogues of ballistic transport equations.
In trying to find such algebras, explicit constructions will begin with the maximal finite-dimensional subalgebra
$\langle X_n, \vec{Y}_n\rangle_{n\in \{0,\pm 1\}}$, where $X_{-1}, \vec{Y}_{-1}$ are time and space translations,
respectively, $X_0$ is the dilatation generator, $\vec{Y}_0$ is the generator of generalised Galilei transformations
and $X_1$, $\vec{Y}_1$ are `special' meta-conformal transformations.

In section~2, we begin by adding some further aspects of the $1D$ meta-conformal case, especially the finite transformations listed in table~\ref{tab1}
and also a generalisation of the representation of the Lie algebra (\ref{commutators}). From this, an ansatz for the $d$-dimensional construction is
extracted and used in section~3 to find the generic form of the generators. Particular attention will be devoted to construct the terms which will
describe how primary scaling operators will transform under meta-conformal transformations. In section~4 we shall concentrate on the special case of
$d=2$ dimensions, where stronger results are found. First, we find that there exist two distinct, non-isomorphic symmetric algebras, which
are distinguished by the values $p=-1$ and $p=\frac{1}{3}$ of a certain parameter which arises in the construction. Second, in the case $p=-1$,
we shall show that the symmetry algebra is infinite-dimensional and isomorphic to the direct sum of {\em three} Virasoro algebras
(without central charge). As shown in table~\ref{tab1}, the corresponding Lie group includes ortho-conformal transformations of the spatial variables
as a sub-group. The time-dependent transformations might be used to generate the temporal evolution of the physical system. Indeed, the co-variant
two-point function is explicitly seen to describe the relaxation towards an  ortho-conformally two-point function, which reflects the meta-conformal
aspects in this Lie group.
On the other hand, in the case $p=\frac{1}{3}$, we only suceeded to find a finite-dimensional Lie group of dynamical symmetries. The form of the
co-variant two-point function is distinct. We conclude in section~5.
An appendix gives the details for finding explicitly the finite transformations from the Lie algebra generators.

%%%%%%%%%%%%%%%%%%%%%%%%%%%%%%%%%%%%%%%%%%%%%%%%%%%%%%%%%%%%%%%%%%%%%%%%%%%%%%%%%%%%%%%%%%%%%%%%%%%%%%%%%%%%%%%%%%%%%%%%%%%%%%%%%
\section{Meta-conformal algebras: general remarks}
%%%%%%%%%%%%%%%%%%%%%%%%%%%%%%%%%%%%%%%%%%%%%%%%%%%%%%%%%%%%%%%%%%%%%%%%%%%%%%%%%%%%%%%%%%%%%%%%%%%%%%%%%%%%%%%%%%%%%%%%%%%%%%%%%

\subsection{Finite $1D$ meta-conformal transformation}

In order to obtain  a better  geometric picture of the meta-conformal transformations (\ref{infinivarconf}),
we begin by deriving the corresponding finite $1D$ meta-conformal transformations. Formally, they are given by the Lie series
$F_Y(\eps,t,r) = e^{\eps Y_{m}} F(0,t,r)$ and $F_X(\eps,t,r) = e^{\eps X_{n}} F(0,t,r)$, with the generators taken from (\ref{infinivarconf}).
They are given as the solutions of the two initial-value problems
\begin{subequations} \label{XYfinit}
\begin{align}
& \Bigl( \partial_{\eps} + (t+\mu r)^{m+1}\partial_r + (m+1)\gamma(t+\mu r)^m\Bigr) F_Y(\eps,t,r) = 0
\label{finitYm}\\
& \Bigl( \partial_{\eps} +t^{n+1}\partial_t+\mu^{-1}
\left[(t+\mu r)^{n+1}-t^{n+1}\right]\partial_r+(n+1)\left(xt^n+\frac{\gamma}{\mu}\left[(t+\mu r)^n-t^n\right]\right)\Bigr) F_X(\eps,t,r) = 0
\label{finitXn}
\end{align}
\end{subequations}
subject to the initial conditions $F_X(0,t,r)=F_Y(0,t,r)=\phi(t,r)$.

Rather than presenting the details of that integration (see the appendix for this), it is more instructive to look immediately at the result,
see also table~\ref{tab1}. A simple
form is obtained by using the variable $\rho=t+\mu r$ instead of $r$:
\begin{subequations} \label{finit-1D}
\begin{align}
Y_m:&\quad \phi'(t,\rho)  =  \left(\frac{\D \rho'}{\D\rho}\right)^{\gamma/\mu}\phi(t',\rho')\;\;;\;\;
         t'=t\;\;,\;\; \rho'  =  a(\rho) \\
X_n:&\quad \phi'(t,\rho) =  \left(\frac{\D t'}{\D t}\right)^{\frac{\gamma}{\mu}-x}
                            \left(\frac{\D \rho'}{\D\rho}\right)^{\frac{\gamma}{\mu}} \phi\left(t',\rho'\right)\;\;;\;\;
t'=\beta(t) \;\;,\;\; \rho'=\beta(\rho)
\end{align}
\end{subequations}
where $a=a(\rho)$ and $\beta=\beta(t)$ are arbitrary functions. By expanding $\beta(t)=t+\eps t^{n+1}$ and $a(\rho)=\rho+\eps \rho^{m+1}$ the differential
equations for the Lie series as they follow from the explicit generators can be recovered.

Eqs.~(\ref{finit-1D}) give the global form of the $1D$  meta-conformal transformations and the transformation of the associated primary scaling operators.
Since this work is mainly interested in finding new meta-conformal symmetries, we shall leave the construction of the full conformal field-theory
based on (\ref{finit-1D}) to future work.

We also note the transformation of $r$ as generated by $X_n$
\BEQ
r'=\frac{1}{\mu}[\beta(t+\mu r)-\beta(t)].\label{rtransf}
\EEQ

\subsection{A generalisation of the one-dimensional case}
Considering (\ref{ineq1}), or else (\ref{Boltzmannd}) restricted to $d=1$, and focussing on the finite-dimensional sub-algebra of symmetries,
it is known that the generators $X_1,Y_{0,1}$ can be generalised as follows, with new constants $\alpha,\beta$ \cite{Stoimenov15}:
\BEA
X_1   & = & -\left(t^2+\alpha r^{2}\right)\partial_t - \left(2t r +\beta r^{2}\right)\partial_r-2x t - 2\mu x r,\nonumber\\
Y_{0} & = & -\alpha r\partial_t-\left(t + \beta r\right)\partial_r -\gamma \nonumber\\
Y_{1} & = & -\alpha\left(2t r+\beta r^{2}\right)\partial_t
\label{extendsymmetries} \\
 && - \left(t^2+2\beta t r +(\alpha+\beta^2)r^{2}\right)\partial_r - 2\gamma t -2\left(\alpha x+\beta\gamma\right)r. \nonumber
\EEA
while the generators $X_{-1,0}, Y_{-1}$ are un-modified with respect to (\ref{infinivarconf}).
For $n,m,\in \{0,\pm 1\}$ they satisfy the following commutation relations
\BEA
     &&[X_n, X_{m}]=(n-m)X_{n+m} \;\; ,\;\; [X_n, Y_m] = (n-m)Y_{n+m}\nonumber\\
     && [Y_n, Y_{m}] = (n-m)\left(\alpha X_{n+m} + \beta Y_{n+m}\right).
     \label{metaconformal}
\EEA
Although these commutators look different, the Lie algebra
$\langle X_n, Y_n\rangle_{n\in \{0,\pm 1\}} \cong \mathfrak{sl}(2,\mathbb{R})\oplus\mathfrak{sl}(2,\mathbb{R})$
is isomorphic to the ortho-conformal algebra \cite[Prop. 1]{Stoimenov15} and hence also to (\ref{commutators}).
Furthermore, the generators (\ref{extendsymmetries}) are indeed dynamical meta-conformal symmetries of the  $1D$ eq.~(\ref{Boltzmannd}),
if the parameters are chosen as follows
%\textcolor{blue}{\tt il me semble que la seconde relation est superflue. D'accord ?}
\BEQ
\alpha=\frac{1+\beta c}{c^2} \;\; , \;\;
%\textcolor{blue}{\alpha+\beta^2=\frac{1+\beta c+\beta^2c^2}{c^2}} \;\; , \;\;
x=-\gamma c \;\; , \;\;
c= -\mu^{-1}
\EEQ
Then the dynamical symmetries follow from the commutators
\BEA
&& [{\hat B}, X_1] = -2\left(t+\frac{1+\beta c}{c^2}cr\right){\hat B}\nonumber\\
&& [{\hat B}, Y_0] = -\frac{1+\beta c}{c^2}c{\hat B} \\
&& [{\hat B}, Y_1] = -2\frac{1+\beta c}{c^2}\left(ct+(1+\beta c)r\right){\hat B}\nonumber
\EEA
such that the solution space of (\ref{Boltzmannd}) is indeed invariant.

In order to write an infinite dimensional representation of the algebra (\ref{metaconformal}) we first obtain the generator $X_2$ from
\BEQ [X_2, X_{-1}]=3X_{1}, \quad [X_2, X_0]=2X_2, \quad [X_2, Y_{-1}]=3Y_1\label{x2system}
\EEQ
Starting from a general form
\BEQ X_2 = -a(t,r)\partial_t-b(t,r)\partial_r-c(t,r),
\nonumber
\EEQ
and satisfying the system (\ref{x2system}) we obtain
\BEA a(t,r) & = & t^3+3\alpha tr^2+\alpha\beta r^3\nonumber\\
     b(t,r) & = & 3t^2r+3\beta tr^2 +(\alpha+\beta^2 )r^3\nonumber\\
     c(t,r) & = & 3xt^2+6\gamma tr +3(\alpha x+\beta\gamma )r^2\nonumber
\EEA
Next, the higher members of $X_n$ hierarchy can be obtained from the commutator
\BEQ X_n = \frac{1}{n-2}[X_{n-1}, X_1].\label{xnhierarchy}
\EEQ
We obtain
\BEA X_2 &=& -\left(t^3+\alpha (3tr^2+\beta r^3)\right)\partial_t-
\left(3t^2r+3\beta tr^2 +\beta^2r^3+\alpha r^3\right)\partial_r\nonumber\\
         & & - 3xt^2-3\gamma (2tr+\beta r^2)-3\alpha x r^2\label{x2}\\
     X_3 &=& -\left(t^4+\alpha (6t^2r^2+4\beta t r^3+\beta^2r^4 )+\alpha^2r^4\right)\partial_t\nonumber\\
         & & -\left(4t^3r+6\beta t^2r^2 +4\beta^2tr^3+\beta^3r^4+2\alpha(2tr^3+\beta r^4)\right)\partial_r\nonumber\\
         & & - 4xt^3-4\gamma (3t^2r+3\beta tr^2+\beta^2 r^3)-4\alpha x(3tr^2+\beta r^3)-4\alpha\gamma r^3\label{x3}\\
     X_4 &=& -\left(t^5+\alpha (10t^3r^2+10\beta t^2r^3+5\beta^2 t r^4+\beta^3r^5)+2\alpha^2(5tr^4+2\beta r^5)\right)\partial_t\nonumber\\
         & & -\left(\frac{1}{\beta}[(t+\beta r)^5-t^5]+\alpha(10t^2r^3+10\beta tr^4+3\beta^2r^5)+\alpha^2r^5\right)\partial_r\nonumber\\
         & & - 5xt^4-5\frac{\gamma}{\beta}[(t+\beta r)^4-t^4]-5\alpha x(6t^2r^2+4\beta tr^3+\beta^2r^4)-4\alpha\gamma(2tr^3+\beta r^4)-5\alpha^2 x r^4.\nonumber\\
          \label{x4}
\EEA
 Correspondingly, the higher members of $Y_n$ hierarchy are obtained by the commutator
 \BEQ
 Y_n=n^{-1}[X_n,Y_0]\label{yn}.
 \EEQ
 We obtain
 \BEA Y_2 &=& -\alpha\left(3t^2r+3\beta tr^2+\beta^2 r^3+\alpha r^3\right)\partial_t-
\left((t+\beta r)^3+\alpha(3tr^2+2\beta r^3)\right)\partial_r\nonumber\\
         & & - 3\gamma(t+\beta r)^2-3\alpha(\beta x+\gamma)r^2\label{y2}\\
     Y_3 &=& -\alpha\left(4t^3r+6\beta t^2r^2+4\beta^2 t r^3+\beta^3r^4+2\alpha(2tr^3+\beta r^4)\right)\partial_t\nonumber\\
         & & -\left((t+\beta r)^4+\alpha(6t^2r^2 +8\beta tr^3+3\beta^2r^4)+\alpha^2r^4\right)\partial_r\nonumber\\
         & & - 4\gamma(t+\beta r)^3-\alpha x\frac{4}{\beta}\left((t+\beta r)^3-t^3\right)-4\alpha\gamma(3tr^2+2\beta r^3)-4\alpha^2 xr^3\label{y3}\\
     Y_4 &=& -\frac{\alpha}{\beta}\left((t+\beta r)^5-t^5+\alpha\left(15\beta t^2r^3+\frac{35}{2}\beta^2 tr^4+\frac{11}{2}\beta^3r^5\right)
             +\frac{9}{4}\alpha^2\beta r^5\right)\partial_t\nonumber\\
         & & -\left((t+\beta r)^5-t^5+\alpha(10t^3r^2+20\beta t^2r^3+15\beta^2tr^4+4\beta^3r^5)+\frac{5}{4}\alpha^2(3tr^4+2\beta r^5)\right)\partial_r\nonumber\\
         & & - 5\gamma(t+\beta r)^4-\alpha x\frac{5}{\beta}\left((t+\beta r)^4-t^4\right)-5\alpha\gamma(6t^2r^2+8\beta tr^3+3\beta^2r^4)\nonumber\\
         & & -5\alpha^2 x(4tr^3+2\beta r^4)-5\alpha^2\gamma r^4.\label{y4}
\EEA
  One can conclude that an infinite-dimensional structure in the generators $X_n, Y_n$ is possible to exist up to terms linear in $\alpha$. However for
  $n\geq 2$ the terms with higher power of $\alpha$ appear although the commutation relation of the algebra are steel valid for $n=3$. In particular
  the strange coefficients in $Y_4$ alarms that the commutation relations are no more satisfied(see for example $[X_{-1},Y_4]= -5Y_3$). It follows that
  an infinite-dimensional representation of meta-conformal algebra exists only for $\alpha=0$ and is given by the generators (\ref{infinivarconf}).

\subsection{Ansatz for the $d$-dimensional case}

Higher-dimensional analogues of the meta-conformal algebra (\ref{commutators}) are sought as dynamical symmetries
of a ballistic transport equation, of the form
\BEQ  \label{Boltzmannd}
{\hat B}f(t,\vec{r})=(\partial_t+\vec{c}\cdot \vec{\partial}_{\vec{r}})f(t,\vec{r})=0
\EEQ
where $\vec{c}\in\mathbb{R}^d$ is a constant vector, which naturally generalizes eq.~(\ref{ineq1}).\\

Then the generators of translations and dynamical scaling are trivially
generalized to the $d$-dimensional case
\begin{subequations}
\begin{align}
     X_{-1}   &=  -\partial_t                                         \label{timetranslations}\\
     Y^j_{-1} &=  -\partial_{r_j}\;\; , \quad j\in \{1,...d\}         \label{spacetranslations}\\
     X_0      &=  -t\partial_t-\vec{r}\cdot\partial_{\vec{r}}-\delta  \label{dynscaling}
\end{align}
\end{subequations}
where $\delta$ stands for a scaling dimension.
Then the form of all generators follows from the one of $X_1$. We make the following ansatz
\BEA
X_1 & := &  -(t^2+\alpha\vec{r}^2)\partial_t-2t\vec{r}\cdot\partial_{\vec{r}}-p\vec{r}^2\vec{\beta}\cdot\partial_{\vec{r}}-
            (1-p)(\vec{\beta}\cdot\vec{r})\vec{r}\cdot\partial_{\vec{r}}-2\delta t\nonumber\\
    &    & -\vec{A}(\vec{r},\vec{\beta})\cdot\partial_{\vec{\beta}}-\vec{B}(\vec{r}, \vec{\beta}, \vec{\gamma})\cdot\partial_{\vec{\gamma}}
           -k\vec{\gamma}\cdot\vec{r}\label{X1ddim}
\EEA
where $\alpha$, $p$ and $k$ are scalars, $\vec{\beta}$, $\vec{\gamma}$ are constant vectors and
$\vec{A}, \vec{B}$ are vector functions of their arguments. All these must be found self-consistently from the algebra we are goign to construct.
The form of the $X_1$ is motivated as follows.
\begin{itemize}
\item Taking into account the form of $X_1$, eq.~(\ref{extendsymmetries}) for $d=1$ space dimension,  we see that the sum of the
prefactors of the terms quadratic in $\vec{r}$ must give unity
%%one understand why the sum of the factors between quadratic in $\vec{r}$ terms must gives $1$($p+1-p=1$).
\item $X_1$ should be invariant under spatial rotations
\BEQ \label{invrot}
[X_1, R_{ij}]=0, \quad R_{ij}= r_i \partial_{r_j} - r_j \partial_{r_i}.
\EEQ
However, even in the simplest case when
\BEQ \label{simplest}
\vec{A}=\vec{B}=\vec{\gamma}=0
\EEQ
the resulting form of $X_1$ eq.~(\ref{X1ddim}) is not rotation-invariant under the $R_{ij}$ (\ref{invrot}).
Therefore, it will be necessary to include rotations of the vectors $\vec{\beta}$ and $\vec{\gamma}$ such that the
rotation generator becomes
\BEQ \label{newrotations}
\bar{R}_{ij} = r_i \partial_{r_j} - r_j \partial_{r_i} +
\gamma_i \partial_{\gamma_j} - \gamma_j \partial_{\gamma_i}+
\beta_i \partial_{\beta_j} - \beta_j \partial_{\beta_i}
\EEQ
Furthermore, rotation-invariance of $X_1$ will require the possibility that $\vec{A}\ne \vec{0}$ and $\vec{B}\ne 0$.
In addition, taking into account the commutation relation of the one-dimensional case (\ref{extendsymmetries}),
especially $[[X_1, Y^j_{-1}], Y_{-1}^j]\sim Y_{-1}^j$, it follows that $\vec{A}, \vec{B}$ can at most be linear in $\vec{r}$.
\end{itemize}
Additional restriction on the forms of $X_1$ come from the requirement that it should act as a dynamical symmetry of eq.~(\ref{Boltzmannd}).
By `dynamical symmetry' we mean the following required commutator \cite{Niederer72}
\BEQ \label{condsimX1}
[{\hat B}, X_1]=\lambda(t,\vec{r}){\hat B}.
\EEQ
which implies that the space of solutions of ${\hat B}\vph=0$ is invariant under the action of $X_1$.
As we shall see, this requirement leads to new relations between $\alpha,p$ and $\vec{\beta}$.
As an example, consider the case $d=3$. From (\ref{condsimX1}) it follows that $\delta=0$ and the following conditions
\begin{subequations} \label{syst}
\begin{align}
1+\beta_xc_x+\frac{1-p}{2}(\beta_yc_y+\beta_zc_z) &= \alpha c_x^2\label{sys1}\\
p\beta_xc_y+\frac{1-p}{2}\beta_yc_x               &= \alpha c_xc_y\label{sys2}\\
p\beta_xc_z+\frac{1-p}{2}\beta_zc_x               &= \alpha c_xc_z\label{sys3}\\
1+\frac{1-p}{2}\beta_xc_x+\beta_yc_y+\frac{1-p}{2}\beta_zc_z &= \alpha c_y^2\label{sys4}\\
p\beta_yc_x+\frac{1-p}{2}\beta_xc_y               &= \alpha c_xc_y\label{sys5}\\
p\beta_yc_z+\frac{1-p}{2}\beta_zc_y               &= \alpha c_zc_z\label{sys6}\\
1+\frac{1-p}{2}(\beta_xc_x+\beta_yc_y)+\beta_zc_z &= \alpha c_z^2\label{sys7}\\
p\beta_zc_x+\frac{1-p}{2}\beta_xc_z               &= \alpha c_xc_z\label{sys8}\\
p\beta_zc_y+\frac{1-p}{2}\beta_yc_z               &= \alpha c_yc_z.\label{sys9}
\end{align}
\end{subequations}
We look for a solution of the above system for $\vec{\beta}\ne \vec{0}$. Straightforward calculations show that
\begin{itemize}
\item The case $p=1$ leads to contradictions between some of the equations in the system. This means
that in this case the generator $X_1$ cannot be a symmetry.
\item For $p\ne 1$ we have following solution of the system (\ref{syst})
\begin{subequations}
\begin{align}
c_j & =  \frac{2}{p-1}\frac{\beta_j}{\vec{\beta}^2}\;\; , \quad j=x,y,z \label{eqsym}   \\
  \alpha & =  \frac{(p+1)(p-1)}{4}\:\vec{\beta}^2                       \label{detalpha}
\end{align}
\end{subequations}
and the condition (\ref{condsimX1}) is satisfied, with $\lambda(t,\vec{r})=-2t-(p+1)(\vec{\beta}\cdot\vec{r})$.
In particular, it follows that $\alpha=0$ is only possible for $p=-1$.
\end{itemize}
In certain case, where a more general form of $X_1$ with $\vec{A}\ne 0$ or $\vec{B}\ne 0$ is needed, the sought symmetries generated by $X_1$ can
become conditional symmetries, that is some auxiliary conditions on the field $\Phi(t,\vec{r},\vec{\beta},\vec{\gamma})$
must be imposed. This comes from the fact that $\vec{A}\ne 0$ and $\vec{B}\ne 0$ are in general linear functions of $\vec{r}$.

%%%%%%%%%%%%%%%%%%%%%%%%%%%%%%%%%%%%%%%%%%%%%%%%%%%%%%%%%%%%%%%%%%%%%%%%%%%%%%%%%%%%%%%%%%%%%%%%%%%%%%%%%%%%%%%%%%%%%%%%%%%%%%%%%
\section{Meta-conformal algebra in $d>2$ spatial dimensions}
%%%%%%%%%%%%%%%%%%%%%%%%%%%%%%%%%%%%%%%%%%%%%%%%%%%%%%%%%%%%%%%%%%%%%%%%%%%%%%%%%%%%%%%%%%%%%%%%%%%%%%%%%%%%%%%%%%%%%%%%%%%%%%%%%

Our preliminary computations suggest a principal difference between the algebras
which generalize the meta-conformal algebra $\mathfrak{mconf}(1,1)$
to $\mathfrak{mconf}(1,d)$ for $d=2$ and for $d>2$.
We shall consider this two cases separately beginning with generalization for $d>2$. In particular
we shall give possible representations of the algebra  $\mathfrak{mconf}(1,3)$.
Starting from the ansatz (\ref{X1ddim}), we obtain\footnote{In the ansatz (\ref{X1ddim})
we first take $\vec{A}=\vec{B}=0$, but if necessary we shall redefine $X_1$.}
\BEA
Y_0^j & = & \demi[X_1,Y_{-1}^j]\nonumber\\
           & = & -\alpha r_j\partial_t-\left(t+\demi(1-p)(\vec{\beta}\cdot\vec{r})\right)\partial_{r_j}-pr_j\vec{\beta}\cdot\partial_{\vec{r}}
     -\demi(1-p)\beta_j\vec{r}\cdot\partial_{\vec{r}}-(k/2)\gamma_j\label{y0j}
\EEA
In particular for $d=3$ case we have
\BEA Y_0^x & = & -\alpha x\partial_t-\left(t+\beta_xx+\frac{1-p}{2}(\beta_yy+\beta_zz)\right)\partial_x \nonumber\\
           &   & -\left(p\beta_yx+\frac{1-p}{2}\beta_xy\right)\partial_y-\left(p\beta_zx+\frac{1-p}{2}\beta_xz\right)\partial_z-(k/2)\gamma_x\nonumber\\
     Y_0^y & = & -\alpha y\partial_t-\left(p\beta_xy+\frac{1-p}{2}\beta_yxy\right)\partial_x-
\left(t+\beta_yy+\frac{1-p}{2}(\beta_xx+\beta_zz)\right)\partial_y\nonumber\\
           &   & -\left(p\beta_zy+\frac{1-p}{2}\beta_yz\right)\partial_z-(k/2)\gamma_y\nonumber\\
     Y_0^z & = & -\alpha z\partial_t-\left(p\beta_xz+\frac{1-p}{2}\beta_zx\right)\partial_x-
     \left(p\beta_yz+\frac{1-p}{2}\beta_zy\right)\partial_y\nonumber\\
           &   & - \left(t+\beta_zz+\frac{1-p}{2}(\beta_yy+\beta_xx)\right)\partial_z-(k/2)\gamma_z\label{y0xyz}
\EEA
When calculating $[Y_0^j,Y_0^i]$ for $i\ne j$ we obtain taking into account the value of $\alpha=\frac{(p-1)(p+1)}{4}\vec{\beta}^2$
\BEA [Y_0^x,Y_0^y] & = & \frac{(3p-1)(p+1)(p-1)}{8}(\beta_yx-\beta_xy)\partial_t\nonumber\\
                   &   & +\left(\frac{(3p-1)(p+1)}{4}(\beta_x\beta_yx-\beta_x^2y)+\frac{(1-p)^2}{4}(\beta_z^2y-\beta_y\beta_zz)\right)\partial_x\nonumber\\
                   &   & +\left(\frac{(3p-1)(p+1)}{4}(\beta_y^2x-\beta_x\beta_yy)-\frac{(1-p)^2}{4}(\beta_z^2x-\beta_x\beta_zz)\right)\partial_y\nonumber\\
                   &   & +p^2(\beta_y\beta_zx-\beta_x\beta_zy)\partial_z. \label{1newcom}
\EEA
It follows than that for the solutions of the equation
\BEQ\label{haracteristiceq}
p^2-(1-p)^2/4=(p+1)(3p-1)/4=0,
\EEQ that is for $p_1=-1$ and $p_2=1/3$ one can write
\BEQ [Y_0^x,Y_0^y] = -p^2(\beta_z^2R_{xy}+\beta_x\beta_zR_{yz}-\beta_y\beta_zR_{xz})\label{y0xy}
\EEQ
Similar calculations shows that for the same values of $p=-1,1/3$ it is fulfilled
\BEA && [Y_0^x,Y_0^z] = p^2(\beta_y\beta_zR_{xy}+\beta_x\beta_yR_{yz}-\beta_y^2R_{xz})\label{y0xz}\\
     && [Y_0^y,Y_0^z] = -p^2(\beta_x\beta_zR_{xy}+\beta_x^2R_{yz}-\beta_x\beta_yR_{xz})\label{y0yz}
\EEA
Looking at the expressions (\ref{y0xy},\ref{y0xz},\ref{y0yz}) we see two problems
\begin{itemize}
\item On the right-hand side the commutators give the generators $R_{ij}$ instead $\bar{R}_{ij}$.
In fact one can add directly $R_{\beta_i\beta_j}$ to $R_{ij}$ on the right-hand side because
\BEA   0 & = & \beta_z^2R_{\beta_x\beta_y}+\beta_x\beta_zR_{\beta_y\beta_z}-\beta_y\beta_zR_{\beta_x\beta_z}\nonumber\\
         & = & \beta_y\beta_zR_{\beta_x\beta_y}+\beta_x\beta_yR_{\beta_y\beta_z}-\beta_y^2R_{\beta_x\beta_z}\nonumber\\
         & = & \beta_x\beta_zR_{\beta_x\beta_y}+\beta_x^2R_{\beta_y\beta_z}-\beta_x\beta_yR_{\beta_x\beta_z}.\label{zerocond}
\EEA
On the other hand $R_{\gamma_i\gamma_j}$ can not be added directly.
It can be obtained working with a more general form of $X_1$ (\ref{X1ddim}) with
$\vec{A}=\vec{0}$ and appropriate form of $\vec{B}(\vec{r},\vec{\beta},\vec{\gamma})\ne \vec{0}$ (we shall do this in the second subsection).
\item The second problem is also in the right-hand sides of (\ref{y0xy},\ref{y0xz},\ref{y0yz}), namely they contain as pre-factors
the components of vector $\beta$, which enter in generator of rotations $\bar{R}_{ij}$
and are considered as variables together with $x,y,z$ and $\gamma_x,\gamma_y,\gamma_z$. In order to obtain a closed algebraic structure,
which contains the generators $ Y_{-1}^{x,y,z},Y_0^{x,y,z}$ and $\bar{R}_{ij}$ we must require
\BEQ R_{\beta_i\beta_j}=\beta_i\partial_{\beta_j}-\beta_j\partial_{\beta_i}=0\label{fixedbeta}
\EEQ
So to satisfy above we must fix $\vec{\beta}$ such that it is characterized by unique scalar non-zero parameter,
that is if more than one of the components of $\vec{\beta}$ are non-zero, they must be equal(up to sign). We shall restrict
to the case $\beta_x=\beta, \beta_y=\beta_z=0$, that is a ballistic transport in $x$ direction.
\end{itemize}

\subsection{Meta-conformal algebra in $d=3$ space dimensions\\ with $\vec{\beta}=(\beta,0,0)$ and $\vec{\gamma}=\vec{0}$}

Because of the fact that $\vec{\gamma}=\vec{0}$, the generators of rotations have the usual form $R_{ij}=r_i\partial_{r_j}-r_j\partial_{r_i}$.
Then for $\alpha = \frac{1}{4}{(p+1)(p-1)}\beta^2$ we obtain
\BEA X_1 & = & -\left(t^2+\alpha(x^2+y^2+z^2)\right)\partial_t-\left(2tx+\beta x^2+\beta p(y^2+z^2)\right)\partial_x\nonumber\\
                &   & -(2t+\frac{1-p}{2}\beta x)y\partial_y-(2t+\frac{1-p}{2}\beta x)z\partial_z-2\delta t.\label{X1d3A1}
\EEA
{}From the above form of $X_1$ we can generate all other generators. Starting from $Y_0^j=\demi[X_1,Y_1^j]$ we obtain
\BEA
  Y^x_0 & = & -\alpha x\partial_t-(t+\beta x)\partial_x-\frac{1-p}{2}\beta\partial_y-\frac{1-p}{2}\beta z\partial_z\label{Y0xA1}\\
{Y}^y_0 & = &  -\alpha y\partial_t-p\beta y\partial_x-(t+\frac{1-p}{2}\beta x)\partial_y\label{Y0yA1}\\
{Y}^z_0 & = &  -\alpha z\partial_t-p\beta z\partial_x-(t+\frac{1-p}{2}\beta x)\partial_z.\label{Y0zA1}
\EEA
We calculate that $[Y_0^x, Y_0^y]=[Y_0^x, Y_0^z]=0$, but
\BEQ
[Y_0^y, Y_0^z] = -p^2\beta^2R_{yz}\label{1ntrivcom0}.
\EEQ
Next we obtain also
\BEA && [Y_0^x,Y_{-1}^x]=\alpha X_{-1}+\beta Y_{-1}^x\label{2ntrivcomxy0ym1A1}\\
     && [Y_0^y,Y_{-1}^y] =[Y_0^z,Y_{-1}^z]=\alpha X_{-1}+p\beta Y_{-1}^x\label{3nontrivcomyzy0ym1A1}\\
     && [Y_0^x,Y_{-1}^y] = [Y_0^y,Y_{-1}^x] = \frac{1-p}{2}\beta Y_{-1}^y\label{4nontrivcomxyy0ym1A1}\\
     && [Y_0^x,Y_{-1}^z] = [Y_0^z,Y_{-1}^x] = \frac{1-p}{2}\beta Y_{-1}^z.\label{5nontrivcomxzy0ym1A1}
\EEA

Furthermore, from the commutators $Y_1^j=[X_1,Y_{0}^j]$ we calculate
\BEA Y_1^x & = & -\alpha\left(2tx+\beta x^2+(1-2p)\beta(y^2+z^2)\right)\partial_t-2\alpha\delta x\nonumber\\
           &   & -\left(t+2\beta tx+\frac{1-p}{2}\beta^2([p+2]x^2+p[y^2+z^2])\right)\partial_x\nonumber\\
           &   & -(1-p)\beta\left(t+\frac{1-p}{2}\beta x\right)\left(y\partial_y+z\partial_z\right)\label{Y1xA1}\\
     Y_1^y & = & -2\alpha(t+p\beta x)y\partial_t-2p\beta\left(t+\frac{p^2+4p-1}{4p}\beta x\right)y\partial_x-2\alpha\delta y\nonumber\\
           &   & -\left(t^2+(1-p)\beta tx+p^2\beta^2(x^2-y^2+z^2)\right)\partial_y+2p^2\beta^2yz\partial_z\label{Y1yA1}\\
     Y_1^z & = & -2\alpha(t+p\beta x)z\partial_t--2p\beta\left(t+\frac{p^2+4p-1}{4p}\beta x\right)z\partial_x-2\alpha\delta z\nonumber\\
           &   & +2p^2\beta^2zy\partial_y-\left(t^2+(1-p)\beta tx+p^2\beta^2(x^2+y^2-z^2)\right)\partial_z\label{Y1zA1}
\EEA
The nonzero commutators are
\BEA && [Y_1^x,Y_{-1}^x]=2(\alpha X_0+\beta Y_{0}^x)                    \label{6nontrivcomxxy1ym1A1}\\
     && [Y_1^y,Y_{-1}^y] =[Y_1^z,Y_{-1}^z]=2(\alpha X_0+p\beta Y_{0}^x) \label{7nontrivcomyyzzy1ym1A1}\\
     && [Y_1^x,Y_{-1}^y] = [Y_1^y,Y_{-1}^x] =(1-p)\beta Y_{0}^y         \label{8nontrivcomxyy1ym1A1}\\
     && [Y_1^x,Y_{-1}^z] = [Y_1^z,Y_{-1}^x] = (1-p)\beta Y_{0}^z        \label{9nontrivcomxzy1ym1A1}\\
     && [Y_1^y,Y_{-1}^z] = 2\beta^2R_{yz}= -[Y_1^z,Y_{-1}^y]            \label{10nontrivcomyzy1ym1A1}
\EEA
\BEA && [Y_1^x,Y_{0}^x]=\alpha X_1+\beta Y_{1}^x\label{11nontrivcomxxy1y0A1}\\
     && [Y_1^y,Y_{0}^y] =[Y_1^z,Y_{01}^z]=\alpha X_1+p\beta Y_1^x\label{12nontrivcomyyzzy1y0A1}\\
     && [Y_1^x,Y_{0}^y] = [Y_1^y,Y_{0}^x] = \frac{1-p}{2}\beta Y_{1}^y\label{13nontrivcomxyy1y0A1}\\
     && [Y_1^x,Y_{0}^z] = [Y_1^z,Y_{0}^x] = \frac{1-p}{2}\beta Y_{1}^z.\label{14nontrivcomxzy1y0A1}
\EEA
in addition to
\BEA && [X_n,Y_m^j]=(n-m)Y_{n+m}^j,\quad n=0,\pm 1, m= 0, \pm 1, j=x,y,z \label{15nontrivcomxnym}\\
     && [Y_m^y, R_{yz}]=Y_m^z, \quad [Y_m^z, R_{yz}]= -Y_m^y, \quad m= 0, \pm 1.\label{16nontrivcomymryzA1}
\EEA
It follows than that the algebra
\BEQ \mathfrak{mconf}(1,3)=\{X_{0,\pm 1}, Y_{0,\pm 1}^{x,y,z}, R_{yz}\}\nonumber
\EEQ
is closed.

\subsection{Meta-conformal algebra in $d=3$ space dimensions with $\vec{\gamma}\ne\vec{0}$}

We first redefine the generator of rotations
\BEQ
R_{yz}\to \bar{R}_{yz} =: y\partial_z-z\partial_y+\gamma_y\partial_{\gamma_z}-\gamma_z\partial_{\gamma_z}\label{gRyz}.
\EEQ
Then we modify $X_1$(\ref{X1d3A1}) by the following ansatz
\BEA X_1 & \to &  X_1+\tilde{X}_1 \nonumber\\
     \tilde{X}_1 & = & -a(\vec{\beta}\cdot\vec{r})\vec{\gamma}\cdot\partial_{\vec{\gamma}}
     -b(\vec{\gamma}\cdot\vec{r})\vec{\beta}\cdot\partial_{\vec{\gamma}}
     -c(\vec{\beta}\cdot\vec{\gamma})\vec{r}\cdot\partial_{\vec{\gamma}}-k(\vec{\gamma},\vec{r})\nonumber\\
                 & = & -\beta\left((a+b+c)x\gamma_x+a(y\gamma_y+z\gamma_z)\right)\partial_{\gamma_x}-k(x\gamma_x+y\gamma_y+z\gamma_z)\nonumber\\
                 &   & -\beta(bx\gamma_y+cy\gamma_x)\partial_{\gamma_y}-\beta(bx\gamma_z+cz\gamma_x)\partial_{\gamma_z}\label{modX1A1}
\EEA
where $a,b,c$ and $k$ are constants to be determined. Next we generate $Y_0^{x,y,z}$ and $Y_1^{x,y,z}$ in usual way. In particular we obtain
\BEA
     && Y_0^x \to Y_0^x+\tilde{Y}_0^x, \quad Y_0^y \to Y_0^y+\tilde{Y}_0^y, \quad Y_0^z \to Y_0^z+\tilde{Y}_0^z\nonumber\\
     && \tilde{Y}_0^x = -(\beta/2)\left((a+b+c)\gamma_x\partial_{\gamma_x}+b(\gamma_y\partial_{\gamma_y}
                        +\gamma_z\partial_{\gamma_z}\right)-(k/2)\gamma_x\label{modY0xA1}\\
     && \tilde{Y}_0^y = -(\beta/2)(a\gamma_y\partial_{\gamma_x}+c\gamma_x\partial_{\gamma_y})-(k/2)\gamma_y\label{modY0yA1}\\
     && \tilde{Y}_0^z = -(\beta/2)(a\gamma_z\partial_{\gamma_x}+c\gamma_x\partial_{\gamma_z})-(k/2)\gamma_z\label{modY0zA1}
\EEA

In order to satisfy the commutation relations relevant to the previous case
($\vec{\beta}=(\beta,0,0), \vec{\gamma}=\vec{0}$)\footnote{Modifying $X_1$ and correspondingly $Y_0^{x,y,z}$ and $Y_1^{x,y,z}$
by additive terms obviously does not change the commutation relations.} we first obtain
$b=a, \quad c=-a$(from $[Y_0^x,Y_0^y]=[\tilde{Y}_0^x,\tilde{Y}_0^y]=0 $), and $a=\pm 2p$ (from $[Y_0^y,Y_0^z]=-p^2\beta^2\bar{R}_{yz}$)
and next the following representations of meta-conformal algebra $\mathfrak{mconf}(1,3)$
\BEA
     X_1 & = & -\left(t^2+\alpha(x^2+y^2+z^2)\right)\partial_t-\left(2tx+\beta x^2+\beta p(y^2+z^2)\right)\partial_x\nonumber\\
         &   & -(2t+\frac{1-p}{2}\beta x)y\partial_y-(2t+\frac{1-p}{2}\beta\beta x)z\partial_z\nonumber\\
         &   & -a\beta(x\gamma_x+y\gamma_y+z\gamma_z)\partial_{\gamma_x}-a\beta(x\gamma_y-y\gamma_x)\partial_{\gamma_y}\nonumber\\
         &   & -a\beta(x\gamma_z-z\gamma_x)\partial_{\gamma_z}-2\delta t-k(x\gamma_x+y\gamma_y+z\gamma_z),\label{X1d3A2}\\
   Y^x_0 & = & -\alpha x\partial_t-(t+\beta x)\partial_x-\frac{1-p}{2}\beta\partial_y-\frac{1-p}{2}\beta z\partial_z\nonumber\\
         &   & -(a\beta/2)(\gamma_x\partial_{\gamma_x}+\gamma_y\partial_{\gamma_y}+\gamma_z\partial_{\gamma_z})-(k/2)\gamma_x,\label{Y0xA2}\\
 {Y}^y_0 & = & -\alpha y\partial_t-p\beta y\partial_x-(t+\frac{1-p}{2}\beta x)\partial_y\nonumber\\
         &   & -(a\beta/2)(\gamma_y\partial_{\gamma_x}-\gamma_x\partial_{\gamma_y})-(k/2)\gamma_y,\label{Y0yA2}\\
 {Y}^z_0 & = & -\alpha z\partial_t-p\beta z\partial_x-(t+\frac{1-p}{2}\beta x)\partial_z\nonumber\\
         &   & -(a\beta/2)(\gamma_z\partial_{\gamma_x}-\gamma_x\partial_{\gamma_z})-(k/2)\gamma_z,\label{Y0zA2}
\EEA
\BEA
     Y_1^x & = & -\alpha\left(2tx+\beta x^2+(1-2p)\beta(y^2+z^2)\right)\partial_t\nonumber\\
           &   & -\left(t+2\beta tx+\frac{1-p}{2}\beta^2([p+2]x^2+p[y^2+z^2])\right)\partial_x\nonumber\\
           &   & -(1-p)\beta\left(t+\frac{1-p}{2}\beta x\right)\left(y\partial_y+z\partial_z\right)\nonumber\\
           &   & -a\beta\left((t+\beta x)\gamma_x+\frac{1-p}{2}\beta(y\gamma_y+z\gamma_z)\right)\partial_{\gamma_x}-a\beta\left((t+\beta x)\gamma_y
                 -\frac{1-p}{2}\beta y\gamma_x\right)\partial_{\gamma_y}\nonumber\\
           &   & -a\beta\left((t+\beta x)\gamma_z-\frac{1-p}{2}\beta z\gamma_x\right)\partial_{\gamma_z}-kt\gamma_x-(2\alpha\delta+\beta\gamma_x)x
                 -k\beta\frac{1-p}{2}(\gamma_yy+\gamma_zz),\nonumber\\
           &   &              \label{Y1xA2}
\EEA
\BEA
     Y_1^y & = & -2\alpha(t+p\beta x)y\partial_t-2p\beta\left(t+\frac{p^2+4p-1}{4p}\beta x\right)y\partial_x\nonumber\\
           &   & -\left(t^2+(1-p)\beta tx+p^2\beta^2(x^2-y^2+z^2)\right)\partial_y+2p^2\beta^2yz\partial_z\nonumber\\
           &   & -a\beta\left(p\beta y\gamma_x+(t+\frac{1-p}{2}\beta x)\gamma_y\right)\partial_{\gamma_x}
           +a\beta\left((t+\frac{1-p}{2}\beta x)-p\beta y\gamma_y-\frac{a\beta}{2}z\gamma_z\right)\partial_{\gamma_y}\nonumber\\
           &   & -a\beta\left(p\beta y\gamma_z-\frac{a\beta}{2}z\gamma_y\right)\partial_{\gamma_z}-kt\gamma_y-k\beta\frac{1-p}{2}\gamma_yx
                 -(2\alpha\delta+kp\beta\gamma_x)y,\nonumber\\
           &   & \label{Y1yA2}
\EEA
\BEA Y_1^z & = & -2\alpha(t+p\beta x)z\partial_t--2p\beta\left(t+\frac{p^2+4p-1}{4p}\beta x\right)z\partial_x\nonumber\\
           &   & +2p^2\beta^2zy\partial_y-\left(t^2+(1-p)\beta tx+p^2\beta^2(x^2+y^2-z^2)\right)\partial_z\nonumber\\
           &   & -a\beta\left(p\beta z\gamma_x+(t+\frac{1-p}{2}\beta x)\gamma_z\right)\partial_{\gamma_x}
           -a\beta\left(p\beta z\gamma_y-\frac{a\beta}{2}y\gamma_z\right)\partial_{\gamma_y}\nonumber\\
           &   & +a\beta\left((t+\frac{1-p}{2}\beta x)\gamma_x-\frac{a\beta}{2}y\gamma_y-p\beta z\gamma_z\right)\partial_{\gamma_z}-kt\gamma_z
                 -k\beta\frac{1-p}{2}\gamma_zx-(2\alpha\delta+kp\beta\gamma_x)z.\nonumber\\
           &   & \label{Y1zA2}
\EEA
Note that the above expressions for the representation of the meta-conformal algebra give in fact four different representations,
parameterized by the admissible values of $p=-1,a=-2,2$ and $p=1/3, a=-2/3,2/3$,
while the value of $k$ is not fixed in the level of commutation relations. \\
{\bf Example}:$ p=-1, a=2, k=2$
\BEA X_1 & = & -t^2\partial_t-\left(2tx+\beta(x^2-y^2-z^2)\right)\partial_x-2(t+\beta x)y\partial_y-2(t+\beta x)z\partial_z\nonumber\\
         &   & -\delta t-2(x\gamma_x-y\gamma_y-z\gamma_z)-2\beta(x\gamma_x-y\gamma_y-z\gamma_z)\partial_{\gamma_x}\nonumber\\
         &   & -2\beta(x\gamma_y-y\gamma_x)\partial_{\gamma_y}-2\beta(x\gamma_z-z\gamma_x)\partial_{\gamma_z}\label{X1d31}\\
   Y^x_0 & = & -(t+\beta x)\partial_x-\beta y\partial_y-\beta z\partial_z -\beta\gamma_x\partial_{\gamma_x}-\beta\gamma_y\partial_{\gamma_y}
               -\beta\gamma_z\partial_{\gamma_z}-\gamma_x\label{Y0x1}\\
 {Y}^y_0 & = & \beta y\partial_x-(t+\beta x)\partial_y-\beta\gamma_y\partial_{\gamma_x}+\beta\gamma_x\partial_{\gamma_y}-\gamma_y\label{Y0y1}\\
 {Y}^z_0 & = & \beta z\partial_x-(t+\beta x)\partial_z-\beta\gamma_z\partial_{\gamma_x}+\beta\gamma_x\partial_{\gamma_z}-\gamma_z.\label{Y0z1}\\
   Y_1^x & = & -\left((t+\beta x)^2-\beta^2(y^2+z^2)\right)\partial_x-2\beta(t+\beta x)y\partial_y-2\beta(t+\beta x)z\partial_z\nonumber\\
         &   & -2\beta\left((t+\beta x)\gamma_x+\beta(y\gamma_y+z\gamma_z)\right)\partial_{\gamma_x}-2\beta\left(t\gamma_y+\beta(x\gamma_y-y\gamma_x)\right)
               \partial_{\gamma_y}\nonumber\\
         &   & -2\beta\left(t\gamma_z+\beta(x\gamma_z-z\gamma_x)\right)\partial_{\gamma_z}-2t\gamma_x-4(x\gamma_x+y\gamma_y+z\gamma_z)\label{Y1x1}\\
     Y_1^y & = & 2\beta(t+\beta x)y\partial_x-\left((t+\beta x)^2+\beta^2(z^2-y^2)\right)\partial_y+2\beta^2yz\partial_z\nonumber\\
           &   & -2\beta\left((t+\beta x)\gamma_y-\beta y\gamma_x\right)\partial_{\gamma_x}+2\beta\left((t+\beta x)\gamma_x
                 +\beta(y\gamma_y-z\gamma_z)\right)\partial_{\gamma_y}\nonumber\\
           &   & +2\beta^2(y\gamma_z+z\gamma_y)\partial_{\gamma_z}-2t\gamma_y-4\beta(x\gamma_y-y\gamma_x)\label{Y1y1}\\
     Y_1^z & = & 2\beta(t+\beta x)z\partial_x +2\beta^2yz\partial_y-\left((t+\beta x)^2+\beta^2(y^2-z^2)\right)\partial_z\nonumber\\
           &   & -2\beta\left((t+\beta x)\gamma_z-\beta z\gamma_x\right)\partial_{\gamma_x}+2\beta^2(z\gamma_y+y\gamma_z)\partial_{\gamma_y}\nonumber\\
         &   & -2\beta\left((t+\beta x)\gamma_x+\beta(y\gamma_z-z\gamma_z)\right)\partial_{\gamma_z}-2t\gamma_z-4\beta(x\gamma_z-z\gamma_x).\label{Y1z0}
\EEA

\subsection{Symmetries}
We shall now verify whether the representations
(\ref{timetranslations}, \ref{spacetranslations}, \ref{dynscaling}, \ref{X1d3A2}, \ref{Y0xA2}, \ref{Y0yA2}, \ref{Y0zA2}, \ref{Y1xA2}, \ref{Y1yA2},
\ref{Y1zA2}, \ref{gRyz}) act as symmetry algebra of an equation in the form (\ref{Boltzmannd},\ref{eqsym})
\BEQ\label{inveqA}
{\hat B}_A\Phi_{\vec{\gamma}}(t,\vec{r})=\left(\partial_t+\frac{2}{\beta(p-1)}\partial_x\right)\Phi_{\vec{\gamma}}(t,\vec{r})=0.
\EEQ
 We calculate
\BEA
&& [{\hat B}_A,X_1] = -\left(2t+(p+1)\beta x\right){\hat B}_A-2\left(\delta +\frac{k}{\beta(p-1)}\gamma_x
                      +\frac{a}{p-1}\vec{\gamma}\cdot\partial_{\vec{\gamma}}\right)\label{cond1sym}\\
&&  [{\hat B}_A,X_0]=-{\hat B}_A, \quad [{\hat B}_A,Y_0^x]=-\frac{p+1}{2}\beta {\hat B}_A\nonumber\\
&& [{\hat B}_A,Y_1^x] = -\frac{\beta(p+1)}{2}\left(2t+(p+1)\beta x\right){\hat B}_A\nonumber\\
&& -(p+1)\beta\left(\delta+\frac{2+(p-1)k}{(p-1)(p+1)\beta}\gamma_x
   + \frac{a}{p-1}\vec{\gamma}\cdot\partial_{\vec{\gamma}}\right)\label{cond2sym}\\
&&[{\hat B}_A,X_{-1}]=[{\hat B}_A,Y_{-1}^{x,y,z}]=[{\hat B}_A,Y_0^{y,z}]=[{\hat B}_A,Y_1^{y,z}]=[{\hat B}_A,\bar{R}_{yz}]=0.\nonumber
\EEA
We conclude the following
\begin{itemize}
\item For $\gamma_x=\gamma_y=\gamma_z=0$ the meta-conformal algebra left invariant the equations (\ref{inveqA}) under condition
that $\delta=0$
\item For $\vec{\gamma}\ne \vec{0}$ and under condition that
$\Phi_{\vec{\gamma}}(t,\vec{r})=\Phi(t,\vec{r})$ that is only representations of
the algebra depend on $\vec{\gamma}$, the meta-conformal algebra left invariant the equations(\ref{inveqA})
for $k=1$ and $\gamma_x=(1-p)\beta\delta$.
\item Finally, if the fields depend on $\vec{\gamma}$, the condition they must satisfy is
\BEQ
\left(\delta +\frac{k}{\beta(p-1)}\gamma_x +
\frac{a}{p-1}\vec{\gamma}\cdot\partial_{\vec{\gamma}}\right)\Phi_{\vec{\gamma}}(t,\vec{r})=0\label{condsym}
\EEQ
In addition, $k=1$ is required. It follows that we have on shell or conditional symmetry algebra of equation (\ref{inveqA}).
\end{itemize}

%%%%%%%%%%%%%%%%%%%%%%%%%%%%%%%%%%%%%%%%%%%%%%%%%%%%%%%%%%%%%%%%%%%%%%%%%%%%%%%%%%%%%%%%%%%%%%%%%%%%%%%%%%%%%%%%%%%%%%%%%%%%%%%%%
\section{Meta-conformal algebra in $d=2$ spatial dimensions}
%%%%%%%%%%%%%%%%%%%%%%%%%%%%%%%%%%%%%%%%%%%%%%%%%%%%%%%%%%%%%%%%%%%%%%%%%%%%%%%%%%%%%%%%%%%%%%%%%%%%%%%%%%%%%%%%%%%%%%%%%%%%%%%%%

Generalisations of the one-dimensional case to $d=2$ space dimensions(with points $(t,x,y)\in\mathbb{R}^3$) can be proceed as follows.
The generators of translations and dynamical scaling read
\begin{subequations} \label{gensimple}
\begin{align}
X_{-1}   &= -\partial_t                                     \label{timetranslations2}\\
Y^x_{-1} &= -\partial_x \;\; , \;\;  Y^y_{-1} = -\partial_y \label{xyspacetranslations}\\
X_0      &= -t\partial_t-x\partial_x-y\partial_y-\delta.    \label{xydynscaling}
\end{align}
\end{subequations}
For this case the form of $X_1$ can be obtained from (\ref{X1ddim}) putting $\vec{A}=\vec{B}=\vec{0}$ and $k=2$.
As we shall show it is the simplest case when
one is able to find a closed algebraic structure. We write
\BEA
     X_1 & = & -\left(t^2+\alpha(x^2+y^2)\right)\partial_t-\left(2tx+\beta_xx^2+(1-p)\beta_yxy+p\beta_xy^2\right)\partial_x\nonumber\\
         &   & -\left(2ty+p\beta_yx^2+(1-p)\beta_xxy+\beta_yy^2\right)\partial_y-2\delta t -2\gamma_xx-2\gamma_yy.\label{xyX1}
\EEA
and the generator of rotations reads
\BEQ
\bar{R}_{xy} =x\partial_y - y \partial_x + \gamma_x \partial_{\gamma_y} - \gamma_y \partial_{\gamma_x}+
\beta_x \partial_{\beta_y} - \beta_y \partial_{\beta_x}.\label{xyrotations}
\EEQ
As seen before, we have $[X_1,R_{xy}] = 0$.
Combination with the translations gives the next two generators, namely
\begin{subequations} \label{gen0}
\begin{align}
     Y_0^x &:= \demi [X_1, Y_{-1}^x]\nonumber\\
           &=  -\alpha x\partial_t-\left(t+\beta_xx+\frac{1-p}{2}\beta_yy\right)\partial_x-
           \left(p\beta_yx+\frac{1-p}{2}\beta_xy\right)\partial_y-\gamma_x\label{py0x}\\
     Y_0^y &:= \demi [X_1, Y_{-1}^y]\nonumber\\
           &=  -\alpha y\partial_t-\left(p\beta_xy+\frac{1-p}{2}\beta_yx\right)\partial_x-
           \left(t+\beta_yy+\frac{1-p}{2}\beta_xx\right)\partial_y-\gamma_y.\label{py0y}
\end{align}
\end{subequations}
Because of (\ref{metaconformal}),  it is necessary to verify whether $Y^x_0$ and $Y^y_0$ commute or not. We calculate
\BEA
     \lefteqn{[Y^x_0, Y^y_0] = \alpha\left(\frac{3p-1}{2}(\beta_yx-\beta_xy)\partial_t+x\partial_y-y\partial_x\right)}\nonumber\\
     && +\frac{p+1}{2}\left(\frac{3p-1}{2}\beta_x\beta_yx-(p\beta^2_x+\frac{1-p}{2}\beta^2_y)y\right)\partial_x
     +\frac{p+1}{2}\left((p\beta^2_y+\frac{1-p}{2}\beta^2_x)x-\frac{3p-1}{2}\beta_x\beta_yy\right). ~~~~~~~\nonumber\\
     && =\frac{(3p-1)(p+1)}{8}\left((p-1)(\beta_x^2+\beta_y^2)(\beta_yx-\beta_xy)\partial_t+2(\beta_x\beta_yx+\beta_x^2y)\partial_x
     +2(\beta_y^2x+\beta_x\beta_yy)\partial_y\right),\label{pdet}
\EEA
where in the second row we have substituted $\alpha = \frac{(p+1)(p-1)}{4}\vec{\beta}^2$ taken from (\ref{detalpha}).
It follows that $[Y^x_0, Y^y_0]=0$ for two cases:
\begin{enumerate}
\item $p=-1$ and consequently $\alpha=0$.
\item $p=1/3$, hence $\alpha=-\frac{2}{9}(\beta_x^2+\beta_y^2)$.
\end{enumerate}
We shall take up these two distinct cases separately.

\subsection{The case $p=-1$}
As mentioned in this case $\alpha=0$, so the generators $X_1,Y^x_0$ and $Y^y_0$ reduce to
\begin{subequations} \label{genaddi}
\begin{align}
  X_1   &= -t^2\partial_t-\left(2tx+\beta_xx^2+2\beta_yxy-\beta_xy^2\right)\partial_x\nonumber\\
        &  ~~~-\left(2ty-\beta_yx^2+2\beta_xxy+\beta_yy^2\right)\partial_y-2\delta t -2\gamma_xx-2\gamma_yy.\label{rightX1}\\
  Y_0^x &=  -(t+\beta_xx+\beta_yy)\partial_x-(\beta_xy-\beta_yx)\partial_y-\gamma_x\label{righty0x}\\
  Y_0^y &= -(\beta_yx-\beta_xy)\partial_x-(t+\beta_yy+\beta_xx)\partial_y-\gamma_y.\label{righty0y}
\end{align}
\end{subequations}
The last two generators $Y^x_1 :=[X_1,Y^x_0]$ and $Y^y_1:=[X_1,Y^y_0]$ become
\begin{subequations} \label{gen1}
\begin{align}
Y^x_1 &= -\left(t^2+2t\beta_xx+2t\beta_yy+(\beta_x^2-\beta_y^2)x^2+4\beta_x\beta_yxy-(\beta_x^2-\beta_y^2)y^2\right)\partial_x\nonumber\\
&  ~~~-\left(2t\beta_xy-2t\beta_yx-2\beta_x\beta_yx^2+2(\beta_x^2-\beta_y^2)xy+2\beta_x\beta_yy^2\right)\nonumber\\
&  ~~~-2\gamma_x(t+\beta_xx+\beta_yy)-2\gamma_y(\beta_xy-\beta_yx)\label{righty1x}\\
      Y^x_1 &=
      -\left(2t\beta_yx-2t\beta_xy+2\beta_x\beta_yx^2-2(\beta_x^2-\beta_y^2)xy-2\beta_x\beta_yy^2\right)\partial_x\nonumber\\
      &  ~~~-\left(t^2+2t\beta_xx+2t\beta_yy+(\beta_x^2-\beta_y^2)x^2+4\beta_x\beta_yxy-(\beta_x^2-\beta_y^2)y^2\right)\partial_y\nonumber\\
      &  ~~~-2\gamma_y(t+\beta_xx+\beta_yy)-2\gamma_x(\beta_yx-\beta_xy).\label{righty1y}
\end{align}
\end{subequations}
It is readily checked that $[Y_1^x,Y_1^y]=[X_1,Y_1^x]=[X_1,Y_1^y]=0$ and finally,
that the generators (\ref{gensimple}, \ref{genaddi}, \ref{gen1}, \ref{xyrotations})
satisfy the following commutation relations, with $n,m\in\{0, \pm 1\}$
\BEA
     && [X_n, X_{m}]=(n-m)X_{n+m},                                                   \nonumber\\
     && [X_n, Y_m^x]=(n-m)Y_{n+m}^x \;\; , \;\;  [X_n, Y_m^y]=(n-m)Y_{n+m}^y,        \nonumber\\
     && [Y_n^x, Y_{m}^y]=[Y_n^y, Y_{m}^x]=(n-m)(\beta_yY_{n+m}^x+\beta_xY_{n+m}^y),  \nonumber\\
     && [Y_n^x, Y_{m}^x]=-[Y_n^y, Y_{m}^y]=(n-m)(\beta_xY_{n+m}^x-\beta_yY_{n+m}^y), \nonumber\\
     && [Y_m^x, \bar{R}_{xy}]=Y_m^y \;\; , \;\;  [Y_m^y, \bar{R}_{xy}]=-Y_m^x
\label{mconfcom}
\EEA
It is turned out that the Lie algebra
\BEQ
\mathfrak{mconf}^A(1,2) := \left\langle X_{0,\pm1}, Y_{0,\pm1}^x, Y_{0,\pm1}^y, R_{xy}\right\rangle\label{algebrad2pm1}
\EEQ
is really closed if $\beta_x=\beta, \beta_y=0$(or if $\beta_x=\pm\beta_y$, see previous section), when $\beta$ is just a parameter.
However all the generators acts as dynamical symmetries of the linear differential equation(even for general $\vec{\beta}=(\beta_x,\beta_y)$)
\BEQ
{\hat B}f(t,x,y)=(\partial_t+c_x\partial_x+c_y\partial_y)f(t,x,y)=0\label{inveqxy}
\EEQ
if
\BEA
     && \gamma_xc_x+\gamma_yc_y+\delta=0\label{cgamma}\\
     && \beta_x=-\frac{c_x}{c_x^2+c_y^2}, \quad \beta_y=-\frac{c_y}{c_x^2+c_y^2}.\label{cbeta}
\EEA
Indeed, one has in general that
\BEQ
      [{\hat B},X_{-1}]=[{\hat B},Y_{-1}^x]=[{\hat B},Y_{-1}^y]=0 \;\; , \;\;
      [{\hat B},X_0]= -{\hat B}.
\EEQ
In addition, under the conditions (\ref{cgamma}, \ref{cbeta}), we also have
\BEQ
      [{\hat B},Y_0^x]=[{\hat B}, Y_0^y]=[{\hat B},Y_{1}^x]=[{\hat B},Y_{1}^y]= 0 \;\; , \;\;
      [{\hat B},X_1]= -2t{\hat B}.\label{symmetries}
\EEQ
which implies the invariance of the solution space.

\subsubsection{Infinite-dimensional extension}

A better understanding of the algebraic structure behind the rather akward set (\ref{mconfcom})
of commutators is obtained by choosing the coordinate axes
such that the vector $\vec{\beta}=(\beta,0)$ of the orientation of the ballistic transport is along the $x$-axis.
This leads a first slight simplification
(in what follows, because of (\ref{cbeta}), we simply have $\beta=\beta_x$ and $\beta_y=0$), in particular, the
generator of rotations now reads
\BEQ
R = R_{xy} = x\partial_y - y\partial_x +\gamma_x\partial_{\gamma_y} - \gamma_y\partial_{\gamma_x}
\EEQ
Considering the commutators between $Y_n^x$ and $Y_n^y$, the peculiar signs arising suggest to go over to new generators
\BEQ
Y_n^{\pm} := \demi \left( Y_n^x \pm \II\, Y_n^y \right)
\EEQ
Then the commutators (\ref{mconfcom}) simplify to (with $n,m\in\{\pm 1,0\}$)
\BEA
\left[ X_n, Y_m^{\pm} \right] &=& (n-m) Y_{n+m}^{\pm} \nonumber \\
\left[ Y_n^{\pm}, Y_m^{\pm} \right] &=& \beta\, (n-m) Y_{n+m}^{\pm} \;\; , \;\; \left[ Y_n^{+}, Y_m^{-} \right] \:=\: 0 \label{eq:metaconf_chiral}\\
\left[ R, Y_n^{\pm} \right] &=& \pm {\II}\, Y_n^{\pm} \nonumber
\EEA
In addition, we go over to complex spatial coordinates $z=x-\II y$ and $\bar{z}=x+\II y$. The generators are then re-expressed as follows
\BEA
X_{-1}     &=& - \partial_t \nonumber \\
Y_{-1}^{+} &=& - \partial \nonumber \\
X_0        &=& -t\partial_t - z\partial - \bar{z}\bar{\partial} -\delta \nonumber \\
Y_0^{+}    &=&  -(t+\beta  z)\partial - \gamma \\
X_1        &=& -t^2\partial_t -(2tz+\beta z^2)\partial - (2t\bar{z}+\beta \bar{z}^2)\bar{\partial} -2\delta t -2\gamma z -2\bar{\gamma}\bar{z} \nonumber \\
Y_1^{+}    &=& -(t+\beta z)^2 \partial -2\gamma (t+\beta z) \nonumber
\EEA
where $\partial=\partial_z$, $\bar{\partial}=\partial_{\bar{z}}$ and the complex components $\gamma :=\demi(\gamma_x+\II\gamma_y)$ and
$\bar{\gamma}:=\demi(\gamma_x -\II \gamma_y)$. The generators $Y_n^{-}$
are obtained from $Y_n^{+}$ by the replacements $z\mapsto \bar{z}$, $\partial \mapsto \bar{\partial}$ and
$\gamma \mapsto \bar{\gamma}$.  Clearly, restricting to points $(t,z)$ or $(t,\bar{z})$,
we recover  the representation (\ref{infinivarconf}) of the meta-conformal algebra in
$d=1$ dimensions, restricted to $n=1,0,1$.

The algebra (\ref{eq:metaconf_chiral}) is identical to the non-local
meta-conformal algebra found recently for the diffusion-limited erosion process
in one spatial dimension \cite{Henkel17a,Henkel17b}. Therefore,
we define the new generators
\BEQ
A_n := X_n - \frac{1}{\beta} Y_{n}^{+} - \frac{1}{\beta} Y_n^{-}
\EEQ
In the basis $\langle A_n, Y_n^{+}, Y_n^{-}\rangle_{n\in\{\pm 1,0\}}$,
the Lie algebra (\ref{eq:metaconf_chiral}) becomes the direct sum
$\mathfrak{sl}(2,\mathbb{R})\oplus\mathfrak{sl}(2,\mathbb{R})\oplus\mathfrak{sl}(2,\mathbb{R})$.
The above definition can now be extended to an infinite-dimensional set of generators, with $n\in\mathbb{Z}$
\BEA
A_n &=& - t^{n+1} \left( \partial_t - \frac{1}{\beta}\,\partial - \frac{1}{\beta}\, \bar{\partial} \right) - (n+1) t^n \left( \delta -\frac{\gamma}{\beta}
        - \frac{\bar{\gamma}}{\beta} \right)
\nonumber \\
Y_n^{+} &=& - (t+\beta z)^{n+1} \partial - (n+1)\gamma (t+\beta z)^n  \label{eq:syminf2d}\\
Y_n^{-} &=& - (t+\beta \bar{z})^{n+1} \bar{\partial} - (n+1)\bar{\gamma} (t+\beta \bar{z})^n \nonumber
\EEA
with the only non-vanishing commutators
\BEQ \label{3.19}
\left[ A_n, A_m \right] = (n-m) A_{n+m} \;\; , \;\; \left[ Y_n^{\pm}, Y_m^{\pm} \right] = \beta\, (n-m) Y_{n+m}^{\pm}
\EEQ
such that the Lie algebra (\ref{3.19}) is isomorphic
$\mathfrak{vect}(S^1)\oplus\mathfrak{vect}(S^1)\oplus\mathfrak{vect}(S^1)$,
the direct sum of three Virasoro algebras without central charge.
Finally, the ballistic operator (\ref{inveqxy}) becomes
${\hat B}=-\partial_t +\frac{1}{\beta}\left(\partial+\bar{\partial}\right)$ and obeys the commutators
\BEQ
\left[ A_n, {\hat B} \right] = (n+1)t^n  {\hat B} - (n+1)n t^{n-1} \wit{\delta} \;\; , \;\; \left[ Y_n^{\pm}, {\hat B} \right] = 0
\EEQ
where $\wit{\delta} := \delta -\frac{\gamma}{\beta} - \frac{\bar{\gamma}}{\beta}$. Summarising, we have proven:

\noindent
{\bf Proposition:} {\it In two spatial dimensions $\vec{r}=(x,y)$, the linear
ballistic transport equation (\ref{Boltzmannd}) can be brought to the form
${\hat B}f(t,x,y) = \left(-\partial_t + \beta^{-1}\partial_x\right) f(t,x,y)=0$, where $\beta$
is a constant. Its maximal dynamical symmetry is infinite-dimensional, spanned by the generators
(\ref{eq:syminf2d}), if only $\wit{\delta} = \delta -\frac{\gamma}{\beta} - \frac{\bar{\gamma}}{\beta}=0$.
Herein, complex coordinates $z=x-\II y$, $\bar{z}=x+\II y$ and the associated derivatives
$\partial=\partial_z$ and $\bar{\partial}=\partial_{\bar{z}}$ are used and $\gamma,\bar{\gamma},\delta$ are constants.
The Lie algebra of dynamical symmetries is given
by (\ref{3.19}) and is isomorphic to the direct sum of three centre-less Virasoro algebras.}

Working with the coordinates $w=t+\beta z$ and $\bar{w}=t+\beta\bar{z}$, we see that the symmetries
generated by $Y_n^{\pm}$ are ortho-conformal in the variables $(w,\bar{w})$, while the action of
the genertors $A_n$ are meta-conformal. This appears to be the first known example which
combines ortho- and meta-conformal transformations into a single symmetry algebra.
If $\wit{\delta}=0$, we actually have a spectrum-generating algebra for ${\hat B}=A_0$.
In spite of the symmetric formulation, the equation of motion (\ref{inveqxy}) contains a bias, since
the transport goes along the axis $x=\demi\left(z+\bar{z}\right)$, if $\beta\ne 0$.

\subsubsection{Finite transformations}

The finite transformations associated with the generators $A_n,Y_n^{+},Y_n^{-}$ with $n\in\mathbb{Z}$ are given by the corresponding Lie series, for
scaling operators which are scalars under spatial rotations.
The final result is simple:
\begin{subequations} \label{finit-2D}
\begin{align}
Y_n^+:&\quad \phi'(t,z,\bar{z}) = \left(\frac{\D z'}{\D z}\right)^{-\gamma/\mu}\phi(t',z',\bar{z}') \;\;;\;\;
                              t'=t\;\;,\;\; z'=a(z)\;\;,\;\; \bar{z}'=\bar{z} \\
Y_n^-:&\quad \phi'(t,z,\bar{z}) = \left(\frac{\D \bar{z}'}{\D \bar{z}}\right)^{-\gamma/\mu}\phi(t,z,\bar{z}') \;\;;\;\;
                              t'=t\;\;,\;\; z'  =  z\;\;,\;\;  \bar{z}' = \bar{a}(\bar{z})\\
A_n:&\quad \phi'(t,w,\bar{w})  =  \left(\frac{\D t'}{\D t}\right)^{\bar{\delta }}\phi\left(t',w',\bar{w}'\right) \;\;;\;\;
t' = k(t)\;\;,\;\;  w'=w,\quad \bar{w'}=\bar{w}.
\end{align}
\end{subequations}
with the coordinates $w=t+\beta z$, $\bar{w}=t+\beta \bar{z}$ and $k=k(t)$, $a=a(z)$, $\bar{a}=\bar{a}(\bar{z})$ are arbitrary functions.
Expandning these according to $k(t)=t+\eps t^{n+1}$, and analogusly for $a(z)$ and $\bar{a}(\bar{z})$,
the explicit differential equations for the Lie series can be recovered. Their direct integration is detailed in the appendix.

Eqs.~(\ref{finit-2D}) clearly show that the relaxational behaviour decribed by the $2D$ meta-conformal symmetry is governed by {\em three} independent
conformal transformations, rather than two as it is the case for $2D$ conformal invariance at the stationary state.

\subsubsection{Two-point function}

A simple application of dynamical symmetries is the computation of covariantly
transforming two-point functions. Non-trivial results can be obtained from so-called
`{\it quasi-primary}' scaling operators $\phi(t,z,\bar{z})$,
which tranform co-variantly under the finite-dimensional sub-algebra $\langle A_{\pm 1,0}, Y_{\pm 1,0}^{\pm}\rangle$.
Because of temporal and spatial translation-invariance, we can directly write
\BEQ
F(t,z,\bar{z}) = \langle \phi_1(t,z,\bar{z}) \phi_2(0,0,0) \rangle
\EEQ
where the brackets indicate a thermodynamic average which will
have to be carried out when such two-point functions are to computed in the context of a specific statistical mechanics model.
Extending the generators (\ref{eq:syminf2d}) to two-body operators, the covariance is then expressed through the Ward identities
$X_0^{[2]} F = X_1^{[2]} F = Y_0^{\pm,[2]} F = Y_1^{\pm,[2]} F = 0$.
Each scaling operator is characterised by three constants $(\wit{\delta},\gamma,\bar{\gamma})$.
Standard calculations (along the well-known lines of ortho- or meta-conformal invariance) then lead to
\BEQ \label{gl:2pointCase1}
F(t,z,\bar{z}) = F_0\,\delta_{\wit{\delta}_1,\wit{\delta}_2} \delta_{\gamma_1,\gamma_2} \delta_{\bar{\gamma}_1,\bar{\gamma}_2}\:
t^{-2\wit{\delta}_1} (t+\beta z)^{-2\gamma_1} (t+\beta \bar{z})^{-2\bar{\gamma}_1}
\EEQ
where $F_0$ is a normalisation constant. This shows a cross-over between an ortho-conformal two-point function
when $t\ll z,\bar{z}$ and a non-trivial scaling form in the opposite case $t\gg z,\bar{z}$.
We illustrate this for scalar quasi-primary scaling operators, where $\gamma_1=\bar{\gamma}_1$
\BEQ
F(t,z,\bar{z}) \sim \left\{
\begin{array}{ll} t^{-2\wit{\delta}_1} \left( z\bar{z} \right)^{-2\gamma_1}               & \mbox{\rm ~~;~ if $t\ll z,\bar{z}$} \\
                  t^{-2{\delta}_1} \exp\left[-2\beta\gamma_1 \frac{z +\bar{z}}{t}\right]  & \mbox{\rm ~~;~ if $t\gg z,\bar{z}$}
\end{array} \right.
\EEQ
If the time-difference is small compared to the spatial distance, the form of the correlator reduces to the one of standard, ortho-conformal invariance.
For increasing time-differences $t$, the behaviour becomes increasingly close to the known one of effectively $1D$ meta-conformal invariance.\footnote{We did not yet carry out the algebraic procedure which should in the  $t\gg z,\bar{z}$ limit produce the non-diverging behaviour
$F \sim t^{-2{\delta}_1} \exp\left[-2\beta\gamma_1 \left|\frac{z +\bar{z}}{t}\right|\right]$, see \cite{Henkel16}.}

\subsection{The case $p=1/3$}
We shall write the generators $X_1,Y_0^x,Y_0^y$ and non-zero commutation
relations for $\vec{\beta}=(\beta,0)$, that is for $\alpha=-(2/9)\beta^2$
\begin{subequations} \label{pgenaddi}
\begin{align}
  X_1   &= -\left(t^2-\frac{2}{9}\beta^2(x^2+y^2)\right)\partial_t-\left(2tx+\beta x^2+\frac{1}{3}\beta y^2\right)\partial_x\nonumber\\
        &  ~~~-\left(2ty+\frac{2}{3}\beta xy\right)\partial_y-2\delta t -2\gamma_xx-2\gamma_yy.\label{pX1}\\
  Y_0^x &=  \frac{2}{9}\beta^2x\partial_t-(t+\beta x)\partial_x-\frac{1}{3}\beta y\partial_y-\gamma_x\label{pY0x}\\
  Y_0^y &= \frac{2}{9}\beta^2y\partial_t-\frac{1}{3}\beta y\partial_x-\left(t+\frac{1}{3}\beta x\right)\partial_y-\gamma_y.\label{pY0y}
\end{align}
\end{subequations}
The last two generators $Y^x_1 :=[X_1,Y^x_0]$ and $Y^y_1:=[X_1,Y^y_0]$ become
\begin{subequations} \label{gen2}
\begin{align}
Y^x_1 &= \frac{2}{9}\beta^2\left(2tx+\beta x^2+\frac{1}{3}\beta y^2\right)\partial_t-\left(t^2+2\beta t x+\frac{7}{9}\beta^2x^2+\frac{1}{9}\beta^2y^2\right)\partial_x\nonumber\\
&  ~~~-\left(\frac{2}{3}\beta ty+\frac{2}{9}\beta^2xy\right)\partial_y-2\gamma_xt-2\left(\beta\gamma_x-\frac{2}{9}\beta^2\delta\right)x
-\frac{2}{3}\beta\gamma_yy\label{py1x}\\
Y^y_1 &= \frac{2}{9}\beta^2\left(2ty+\frac{2}{3}\beta xy\right)\partial_t-\left(\frac{2}{3}\beta t y+\frac{2}{9}\beta^2xy\right)\partial_x\nonumber\\
&  ~~~-\left(t^2+\frac{2}{3}\beta t x+\frac{1}{9}\beta^2(x^2-y^2)\right)\partial_y-2\gamma_yt-\frac{2}{3}\beta\gamma_yx
-2\left(\frac{1}{3}\beta\gamma_x-\frac{2}{9}\beta^2\delta\right)y\label{py1y}
\end{align}
\end{subequations}
It is readily checked that $[Y_1^x,Y_1^y]=[X_1,Y_1^x]=[X_1,Y_1^y]=0$ and finally,
that the generators (\ref{gensimple}, \ref{genaddi}, \ref{gen1}, \ref{xyrotations})
satisfy the following commutation relations, with $n,m\in\{0, \pm 1\}$
\BEA
     && [X_n, X_{m}]=(n-m)X_{n+m}, \nonumber\\
     && [X_n, Y_m^x]=(n-m)Y_{n+m}^x \;\; , \;\;  [X_n, Y_m^y]=(n-m)Y_{n+m}^y, \nonumber\\
     && [Y_n^x, Y_{m}^y]=[Y_n^y, Y_{m}^x]=\frac{n-m}{3}\beta Y_{n+m}^y, \nonumber\\
     && [Y_n^x, Y_{m}^x]=(n-m)\left(-\frac{2}{9}\beta^2X_{n+m}+\beta Y_{n+m}^x\right), \nonumber\\
     && [Y_n^y, Y_{m}^y]=(n-m)\left(-\frac{2}{9}\beta^2X_{n+m}+\frac{\beta}{3}Y_{n+m}^x\right), \nonumber\\
     && [Y_m^x, \bar{R}_{xy}]=Y_m^y \;\; , \;\;  [Y_m^y, \bar{R}_{xy}]=-Y_m^x
\label{mconfcom2}
\EEA
In addition the Lie algebra
\BEQ
\mathfrak{mconf}^B := \left\langle X_{0,\pm1}, Y_{0,\pm1}^x, Y_{0,\pm1}^y, R_{xy}\right\rangle\label{algebrad2p1over3}
\EEQ
acts as dynamical symmetry algebra of the linear differential equation
\BEQ
{\hat B}_pf(t,x,y)=\left(\partial_t-\frac{3}{\beta}\partial_x\right)f(t,x,y)=0\label{pinveqxy}
\EEQ
if $\delta=(3/\beta)\gamma_x$.\\

Indeed, one has in general that
\BEQ
      [{\hat B}_p,X_{-1}]=[{\hat B}_p,Y_{-1}^x]=[{\hat B}_p,Y_{-1}^y]=0 \;\; , \;\;
      [{\hat B},X_0]= -{\hat B}.
\EEQ
In addition, we obtain
\BEA
     && [{\hat B}_p,Y_0^y]=[{\hat B}_p, Y_1^y]=0 \nonumber\\
     && [{\hat B}_p,X_1]= -\left(2t+\frac{4}{3}\beta x\right){\hat B}_p\nonumber\\
     && [{\hat B}_p, Y_0^x] = -\frac{2}{3}\beta{\hat B}_p\nonumber\\
     && [{\hat B}_p,Y_1^x]= -\frac{4}{3}\beta\left(2t+\frac{2}{3}\beta x\right){\hat B}_p.\label{psymmetries}
\EEA
which proves the symmetries.\\

Looking at the structure of the algebra (\ref{algebrad2p1over3}) it can be shown that
\BEQ
\mathfrak{mconf}^B(1,2)\ncong\mathfrak{sl}(2,\mathbb{R})\oplus\mathfrak{sl}(2,\mathbb{R})\oplus\mathfrak{sl}(2,\mathbb{R}).\nonumber
\EEQ

\subsubsection{Two-point function}

For this case, the two-point function is build by the {\it quasi-primary fields} $\phi(t,x,y,\gamma_x=\gamma,\gamma_y=\lambda)$
which transform covariantly under the algebra $\mathfrak{mconf}^B(1,2)$(\ref{algebrad2p1over3}). Taking into the account the covariance
of time and space translations we write
\BEQ
F=\langle\phi(t_1,x_1,y_1,\gamma_1,\lambda_1)\phi(t_2,x_2,y_2,\gamma_2,\lambda_2)\rangle=F(t,x,y,\gamma_1,\gamma_2,\lambda_1,\lambda_2),\nonumber
\EEQ
where $t=t_1-t_2, x=x_1-x_2, y=y_1-y_2$. The covariance under the other generators is expressed by the following system, to be satisfied by the $F$
\BEA
  X_0:&& (t\partial_t+x\partial_x+y\partial_y+\delta_1+\delta_2)F=0\nonumber\\
Y_0^x:&& \left(\frac{2}{9}\beta^2x\partial_t-(t+\beta x)\partial_x-\frac{\beta}{3}y\partial_y-\gamma_1-\gamma_2\right)F=0\nonumber\\
Y_0^y:&& \left(\frac{2}{9}\beta^2y\partial_t-\frac{\beta}{3}y\partial_x-\left(t+\frac{\beta}{3}x\right)\partial_y-\lambda_1-\lambda_2\right)F=0\nonumber\\
  X_1:&& \left(-\left(t^2+\frac{2}{9}\beta^2x^2+\frac{2}{9}\beta^2y^2\right)\partial_t+\beta\left(x^2+\frac{y^2}{3}\right)\partial_x+\frac{2}{3}\beta xy\partial_y\right)F\nonumber\\
      && -\left(2t\delta_2-2x\gamma_1-2y\lambda_1\right)F=0\nonumber\\
Y_1^x:&& \left(\frac{2}{9}\beta^3\left(x^2+\frac{y^2}{3}\right)\partial_t+\left(t^2-\frac{7}{9}\beta^2x^2-\frac{1}{9}\beta^2y^2\right)
\partial_x-\frac{2}{9}\beta^2xy\partial_y\right)F\nonumber\\
      && +\left(2t\gamma_2
+\beta\left(\frac{4}{9}\beta\delta_1-2\gamma_1\right)x-\frac{2}{3}\lambda_1y\right)F=0\nonumber\\
Y_1^y:&& \left(\frac{4}{27}xy\partial_t-\frac{2}{9}\beta^2xy\partial_x+\left(t^2-\frac{\beta^2}{9}(x^2-y^2)\right)\partial_y\right)F\nonumber\\
      && +\left(2t\lambda_2-\frac{2}{3}\beta\lambda_1x+\beta\left(\frac{4}{9}\beta\delta_1-\frac{2}{3}\lambda_1\right)y\right)F=0\nonumber\label{systemf}\\
R_{xy}:&& \left(x\partial_y-y\partial_x+\gamma_1\partial_{\lambda_1}-\lambda_1\partial_{\gamma_1}+\gamma_2\partial_{\lambda_2}-
\lambda_2\partial_{\gamma_2}\right)F=0\nonumber
\EEA
Next the above system is reduced as follows: the equation corresponding to $X_1$ is replaced by the one corresponding to $\bar{X_1}=X_1+tX_0+xY_0^x+
Y_0^y$ that is instead of fourth equation we write the sum of the forth equation with the first multiplied by $t$,
second multiplied by $x$ and third multiplied
by $y$. It becomes trivial and we retain
\BEQ \label{1eq2}
\gamma_1=\gamma_2=\gamma, \quad \lambda_1=\lambda_2=\lambda, \quad \delta_1=\delta_2=\delta,\nonumber
\EEQ
and the result is substituted in the other equations. In the same way we replace the fifth equation by an equation corresponding to
$\bar{Y}_1^x=Y_1^x-\beta xY_0^x-(\beta/3)yY_0^y$ and the sixth equation by
an equation corresponding to  $\bar{Y}_1^y=Y_1^y-(\beta/3)yY_0^x-(\beta/3)xY_0^y$. We obtain a reduced system acting on
$F=F_1(t,x,y,\gamma,\lambda)$
\BEA
&& (t\partial_t+x\partial_x+y\partial_y+2\delta)F_1=0\nonumber\\
&& \left(\frac{2}{9}\beta^2x\partial_t-(t+\beta x)\partial_x-\frac{\beta}{3}y\partial_y-2\gamma\right)F_1=0\nonumber\\
&& \left(\frac{2}{9}\beta^2y\partial_t-\frac{\beta}{3}y\partial_x-\left(t+\frac{\beta}{3}x\right)\partial_y-2\lambda\right)F_1=0\nonumber\\
&& \left(\left(t^2+\beta t x+\frac{2}{9}\beta^2x^2\right)\partial_x+\frac{\beta}{3}\left(t y+\frac{2}{3}\beta x y\right)\partial_y+2t\gamma
         +\frac{4}{9}\beta^2\delta x\right)F_1=0\nonumber\\
&& \left(\frac{\beta}{3}\left(t y+\frac{2}{3}\beta x y\right)\partial_x+\left(t^2+\frac{2}{3}\beta tx+\frac{2}{9}\beta^2y^2\right)\partial_y
      +2t\lambda+\frac{4}{9}\beta^2y\right)F_1=0\label{redsystemf}\\
&& \left(x\partial_y-y\partial_x+\gamma\partial_{\lambda}-\lambda\partial_{\gamma}\right)F=0\nonumber
\EEA
We shall solve the first three equations, and then substitute the result in the last tree equations, to verify that the system(\ref{redsystemf})
is satisfied.\\

{}From the first equation we express $y\partial_yF_1=-(t\partial_t+x\partial_x+2\delta)F_1$ and put it in the second one. We obtain
\BEQ
\left(\left(t+\frac{2}{3}\beta t x\right)\left(\frac{\beta}{3}\partial_t-\partial_x\right)+\frac{2\beta}{3}
\left(\delta-\frac{3\gamma}{\beta}\right)\right)F=0
.\label{2eq1}
\EEQ
Then we perform a change of variables $F_1(t,x,y)\to G(t,u=t+(2/3)\beta x,y); \partial_t\to \partial_t+\partial_u, \partial_x\to (2/3)\beta\partial_u$.
The equation becomes
\BEQ
\left(\partial_t-\partial_u+\frac{2}{u}\left(\delta-\frac{3\gamma}{\beta}\right)\right)G=0.\label{2eq2}
\EEQ
Another change of variables $G(t,u,y)\to H(t,v=t+u,y);\partial_t\to \partial_t+\partial_v, \partial_u\to \partial_v$ allows us to
obtain
\BEQ
\left(\partial_t+\frac{2}{v-t}\left(\delta-\frac{3\gamma}{\beta}\right)\right)H(t,v,y)=0\label{2eq3}
\EEQ
and determine the dependence on $t$ in $H(t,v,y)$(integrating with respect to $v-t$) namely
\BEQ  \label{1result}
H(t,v,y)=H_0(v,y)(v-t)^{2(\delta-3\gamma/\beta)}.
\EEQ
Next applying the same chain of change of variables we can bring the third equation in the system(\ref{redsystemf})
to the form
\BEA && \left(\frac{2}{9}\beta^2y(\partial_t+\partial_v)-\demi v\partial_y-2\lambda\right)H(t,v,y)\nonumber\\
     && = (v-t)^{2(\delta-3\gamma/\beta)}\left(\frac{2}{9}\beta^2y\partial_v-\demi v\partial_y-2\lambda\right)H_0(v,y)=0.\label{3eq2}
\EEA
The general solution of the above equation is given by the product of a solution of inhomogeneous equation and general solution
of homogeneous one. We obtain
\BEQ H_0(v,y)=H_{01}(w)\exp\left(-\frac{2\lambda}{3\beta}\arctan\left(\frac{\beta|y|}{3|t+\beta x/3|}\right)\right) \quad
w=\left(t+\frac{\beta}{3}x\right)^2+\frac{\beta^2}{9}y^2.\label{2result}
\EEQ
Now substituting result (\ref{3result}) in the first equation of (\ref{redsystemf}) we obtain
\BEQ
wH'_{01}(w)+(2\delta-3\gamma/\beta)H_{01}(w)=0, \quad H_{01}(w)=H_{00}w^{3\gamma/\beta-2\delta}.\label{3result}
\EEQ
Taking the expression (\ref{1result}, \ref{2result}, \ref{3result}) we conclude that the solution of the system (\ref{redsystemf})
and correspondingly the form of two-point function covariant under the algebra $\mathfrak{mconf}^B(1,2)$ is the following

\BEA
F(t,x,y) & = & F_0\,\delta_{\delta_1,\delta_2}\delta_{\gamma_1,\gamma_2} \delta_{\lambda_1,\lambda_2}\left(\left(t+\frac{\beta}{3}x\right)^2
+\frac{\beta^2}{9}y^2\right)^{3\gamma/\beta-2\delta}
\left(t+\frac{2}{3}\beta x\right)^{2(\delta-3\gamma/\beta)}\times\nonumber\\
         & \times & \exp\left(-\frac{2\lambda}{3\beta}\arctan\left(\frac{\beta|y|}{3|t+\beta x/3|}\right)\right).\label{endresult}
\EEA

%%%%%%%%%%%%%%%%%%%%%%%%%%%%%%%%%%%%%%%%%%%%%%%%%%%%%%%%%%%%%%%%%%%%%%%%%%%%%%%%%%%%%%%%%%%%%%%%%%%%%%%%%%%%%%%%%%%%%%%%%%%%%%%%%
\section{Conclusions}
%%%%%%%%%%%%%%%%%%%%%%%%%%%%%%%%%%%%%%%%%%%%%%%%%%%%%%%%%%%%%%%%%%%%%%%%%%%%%%%%%%%%%%%%%%%%%%%%%%%%%%%%%%%%%%%%%%%%%%%%%%%%%%%%%

In this work, we have explored possible mathematical symmetries, using the linear ballistic transport equation as a simple
starting point. Our main results are collected in table~\ref{tab1}: The linear transport equation (\ref{ineq1})
admits in $1D$ and in $2D$ infinite-dimensional Lie groups of dynamical symmetries. In $2D$, these symmetries contain
ortho-conformal transformations of the spatial variables as a sub-group and describe the relaxation of the two-time correlator
towards it. It will be possible to adapt constructions from local scale-invariance \cite{Henkel10,Henkel17c} such that this
Lie group can also describe the relaxation of single-time correlators, which is work in progress.
Remarkably, the Lie algebra of dynamical symmetries in $2D$
is isomorphic to the one of the spatially non-local stochastic process of $1D$ diffusion-limited erosion \cite{Henkel17b}.

Furthermore, the $2D$ case actually admits two distinct, non-isomorphic symmetries. The meaning of these two distinct symmetries remains to
be understood.

Given the considerable variety of strongly interacting systems with dynamical exponent $z=1$, explicit model studies are required
in order to see which of the several kinds of conformal invariance will describe one or several of the known physical situations.
Here a comparison with the explicit correlators (\ref{gl:2pointCase1},\ref{endresult}) might be instructive.

The other important question is to see how to set up the analogue of the ortho-conformal bootstrap.
Work along these lines is in progress. \\

\noindent
{\bf  Ackowledgements:} We warmly thank the organisers of the 10$^{\rm th}$ International Symposium ``Quantum Theory and Symmetries''
and of the atelier ``Lie Theory and Applications in Physics XII'' in Varna (June 2017)
for the excellent atmosphere, where a large part of this work was done.
This work was supported by PHC Rila and by Bulgarian National Science Fund Grant DFNI - T02/6.
.

\newpage
%%%%%%%%%%%%%%%%%%%%%%%%%%%%%%%%%%%%%%%%%%%%%%%%%%%%%%%%%%%%%%%%%%%%%%%%%%%%%%%%%%%%%%%%%%%%%%%%%%%%%%%%%%%%%%%%%%%%%%%%%%%%%%%%%
\appsektion{Finite meta-conformal transformations}
%%%%%%%%%%%%%%%%%%%%%%%%%%%%%%%%%%%%%%%%%%%%%%%%%%%%%%%%%%%%%%%%%%%%%%%%%%%%%%%%%%%%%%%%%%%%%%%%%%%%%%%%%%%%%%%%%%%%%%%%%%%%%%%%%

We provide the details for the explicit integration of the Lie series,
in order to construct the non-infinitesimal, finite meta-conformal transformations. The results are included in table~\ref{tab1}.

\subsection{One spatial dimension}

The Lie series $F_Y(\eps,t,r) = e^{\eps Y_{m}} F(0,t,r)$ and $F_X(\eps,t,r) = e^{\eps X_{n}} F(0,t,r)$ are solutions of the two initial-value problems
eqs.~(\ref{finitYm},\ref{finitXn}),
subject to the initial conditions $F_X(0,t,r)=F_Y(0,t,r)=\phi(t,r)$. It turns out, however, that the calculations are simplified with the
new coordinate $\rho:=t+\mu r$.

We begin by finding $F_Y=F(\eps,t,\rho)$. The initial-value problem (\ref{finitYm}) simplifies to
\BEQ
\Bigl( \partial_{\eps} + \mu\rho^{m+1}\partial_{\rho}-(m+1)\gamma\rho^m\Bigr) F(\eps,t,r) = 0 \;\; , \;\; F(0,t,\rho)=\phi(t,\rho)
\label{1finitym}
\EEQ
Following \cite{Kamke59}, the change of variables $F(\eps,t,\rho)= G(\eps,t,v)$, where $v=\eps+1/(m\mu\rho^m)$, reduces this to
\BEQ
\left(\partial_{\eps}+\frac{(m+1)}{m}\frac{\gamma}{\mu}\frac{1}{v-\eps}\right)G(\eps,t,v)=0\nonumber
\EEQ
and integration yields
\BEQ
G(\eps,t,v)=H(t,v)(v-\eps )^{\frac{\gamma}{\mu}\frac{m+1}{\mu}}.
\label{1resym}
\EEQ
Therein, the initial condition $G(0,t,u)=\phi(t,r)$ fixes the last undetermined function. Setting $\eps=0$, we find
$H(t,\bar{v})={\bar{v}}^{-\frac{\gamma(m+1)}{\mu m}}\phi(t,(n\mu v)^{-1/n})$. Finally, (\ref{1resym}) becomes
\begin{subequations}
\begin{align}
F(\eps,t,\rho) \:=\: \phi'(t,\rho) & =  \left(\frac{\D a(\rho)}{\D\rho}\right)^{-\gamma/\mu}\phi(t',\rho')        \label{endresym}   \\
         t'=t \;\;,\;\; \rho' & =  a(\rho)=\frac{\rho}{[1+\eps \mu m\rho^m]^{1/m}}\label{ymtrcoord}
\end{align}
\end{subequations}
where in the second line we give the corresponding transformation of the coordinates.\\

The second initial value problem (\ref{finitXn}) for $F_X=F(\eps,t,\rho)$ becomes
\BEQ
\Bigl( \partial_{\eps} +t^{n+1}\partial_t+\rho^{n+1}\partial_{\rho}
+(n+1)\left(xt^n+\frac{\gamma}{\mu}\left[\rho^n-t^n\right]\right)\Bigr) F(\eps,t,\rho) = 0 \;\; , \;\;
F(0,t,\rho) = \phi(t,\rho)
\EEQ
With the change of variables $F(\eps,t,\rho)=G(\eps,u,v)$, where $u=\eps+1/(n\mu t^n)$ and $v=\eps+1/(n\mu\rho^n)$, this becomes
\BEQ
\left(\partial_{\eps}+\frac{n+1}{n}(\frac{\gamma}{\mu}-x)\frac{1}{u-\eps}+\frac{n+1}{n}\frac{\gamma}{\mu}\frac{1}{u-\eps}\right)G(\eps,u,v)=0\nonumber
\EEQ
and integrating with respect to $\eps$, we find
\BEQ
G(\eps,u,v)=H(u,v)(u-\eps )^{\frac{n+1}{n}(\frac{\gamma}{\mu}-x)}(v-\eps )^{\frac{n+1}{n}\frac{\gamma}{\mu}}.
\label{xn1res}
\EEQ
Satisfying the initial condition $G(0,u,v)=\phi(t,\rho)$ from where we obtain
first for $\eps=0$
\BEQ
H(\bar{u},\bar{v})=u^{-\frac{n+1}{n}(\frac{\gamma}{\mu}-x)}v^{-\frac{n+1}{n}\frac{\gamma}{\mu}}\phi\left((n\bar{u})^{-1/n},(n\bar{v})^{-1/n}\right)
\nonumber\\
H(u,v)=u^{-\frac{n+1}{n}(\frac{\gamma}{\mu}-x)}v^{-\frac{n+1}{n}\frac{\gamma}{\mu}}\phi\left((n\bar{u})^{-1/n},(n\bar{v})^{-1/n}\right)
\nonumber
\EEQ
Finally substituting in (\ref{xn1res}) for the solution of the second initial problem we find
\BEA
G(\eps,u,v)=\phi'(t,r) & = &\left(1+\eps n t^n\right)^{-\frac{n+1}{n}(\frac{\gamma}{\mu}-x)}
\left(1+\eps n \rho^n\right)^{-\frac{n+1}{n}\frac{\gamma}{\mu}}
\phi\left(\frac{t}{(1+\eps n t^n)^{1/n}}, \frac{\rho}{(1+\eps n \rho^n)^{1/n}}\right)\nonumber\\
                      & = & \left(\frac{\D\beta(t)}{\D t}\right)^{\frac{\gamma}{\mu}-x}
                            \left(\frac{\D\beta(\rho)}{\D\rho}\right)^{\frac{\gamma}{\mu}}\phi\left(t',\rho'\right)
\EEA
where we have defined $\beta(z)=\frac{z}{(1+\eps n z^n)^{1/n}}$. The coordinate transformations now simply read
\BEQ
t'=\beta(t) \;\;,\;\; \rho'=\beta(\rho).
\EEQ

\subsection{Two spatial dimensions, case $p=-1$}

In two spatial dimensions, the meta-conformal algebra with $p=-1$ is spanned by the generators $\langle A_n, Y_n^{+}, Y_n^{-}\rangle_{n\in\mathbb{Z}}$.
We look for the Lie series $e^{\eps A_n}$, $e^{\eps Y_n^{+}}$ and $e^{\eps Y_n^{-}}$.

Since the generator $Y_n^+(Y_n^-)$ can be viewed as a generator meta-conformal algebra
in one spatial dimension, taking into account the expressions (\ref{endresym}, \ref{ymtrcoord})
we can directly write the finite transformations generated by the generators $Y_n^+$ and $Y_n^-$
\begin{subequations} \label{fintrynplus}
\begin{align}
Y_n^+:\quad \phi'(t,z,\bar{z}) & =  \left(\frac{\D a(z)}{\D z}\right)^{-\gamma/\mu}\phi(t',z',\bar{z}),\\
                             t'=t \;\; , \;\; z' & =  a(z)=z(1+\eps \mu n\rho^n)^{-1/n} \;\; , \;\; \bar{z}'=\bar{z}.\label{ynpluscoord}
\end{align}
\end{subequations}
\begin{subequations} \label{finitynminus}
\begin{align}
Y_n^-:\quad \phi'(t,z,\bar{z}) & =  \left(\frac{\D a(\bar{z})}{\D \bar{z}}\right)^{-\gamma/\mu}\phi(t',z,z')),\\
                             t'=t \;\; ,\;\; z' & =  z \;\;,\;\; \bar{z}' = a(\bar{z})=\bar{z}(1+\eps \mu n\bar{z}^n)^{-1/n}.\label{ynminuscoord}
\end{align}
\end{subequations}
The finite transformation generated by $A_n$ are obtained from the following initial value problem
\BEQ
\Bigl( \partial_{\eps} +t^{n+1}\partial_t-\frac{t^{n+1}}{\beta}\partial_{z}
-\frac{t^{n+1}}{\beta}\partial_{\bar{z}}+(n+1)\bar{\delta}t^n\Bigr) F(\eps,t,z,\bar{z}) = 0 \;\;,\;\;
F(0,t,z,\bar{z}) = \phi(t,z,\bar{z})
\label{Anfinit}
\EEQ
The change of variables $F(\eps,t,z,\bar{z})= G(\eps,t,w,\bar{w})$, where $w=t+\beta z,\bar{w}=t+\beta\bar{z}$ cast this into the form
\BEQ
(\partial_{\eps}+t^{n+1}\partial_t+(n+1)\bar{\delta}t^n)G(\eps,t,w,\bar{w})=0 \;\; , \;\;
G(0,t,w,\bar{w})=\phi(t,w,\bar{w}).\label{reducedproblem}
\EEQ
Another change of variables $G(\eps,t,w,\bar{w})= H(\eps,u,w,\bar{w})$ with $u=\eps+t^{-n}/n$ allows us to write instead
\BEQ
\left(\partial_{\eps}+\frac{n+1}{n}\frac{\bar{\delta}}{u-\eps}\right)H(\eps,u,w,\bar{w})=0\nonumber
\EEQ
which is immediately integrated, with the result
\BEQ
H(\eps,u,w,\bar{w})=H_0(u,w,\bar{w})(u-\eps)^{\frac{n+1}{n}\bar{\delta}}.\label{hgeneral}
\EEQ
The initial condition $H(0,u,w,\bar{w})=\phi(t,w,\bar{w})$  fixes the last function $H+0$
\BEQ
H_0(\bar{u},w,\bar{w})=\bar{u}^{-\bar{\delta }\frac{n+1}{n}}\phi\left((n\bar{u})^{-1/n},w,\bar{w}\right)\nonumber
\EEQ
%and than for $\eps\ne 0$
%\BEQ
%H_0(u,w,\bar{w})=(\eps+t^{-n}/n)^{-\frac{n+1}{n}\bar{\delta }}\phi\left(t(1+\eps nt^n)^{-1/n},w,\bar{w}\right).\nonumber
%\EEQ
and the final result reads
\BEQ
H(\eps,u,w,\bar{w})=\phi'(t,w,\bar{w})=(1+\eps nt^n)^{-\frac{n+1}{n}\bar{\delta }}\phi\left(t(1+\eps nt^n)^{-1/n},w,\bar{w}\right).\label{anfinal}
\EEQ
Summarising, this can can rewritten as follows
\begin{subequations} \label{endanfinit}
\begin{align}
\phi'(t,w,\bar{w}) &= \left(\frac{\D k(t)}{\D t}\right)^{\bar{\delta }}\phi\left(t',w',\bar{w}'\right)\\
t' &= k(t)=t(1+\eps nt^n)^{-1/n} \;\;,\;\;  w'=w \;\;,\;\; \bar{w'}=\bar{w}.\label{finalcoordtran}
\end{align}
\end{subequations}
The coordinate transformations in terms of $t,z,\bar{z}$ are easy obtained from the above and read
\BEQ
t'=k(t)\;\;,\;\;z'=z+\frac{t-k(t)}{\beta}\;\;,\;\; \bar{z}'=\bar{z}+\frac{t-k(t)}{\beta }
\EEQ
These take the form of conformal transformations in time and time-dependent translations
in the spatial coordinates $z,\bar{z}$.

%%%%%%%%%%%%%%%%%%%%%%%%%%%%%%%%%%%%%%%%%%%%%%%%%%%%%%%%%%%%%%%%%%%%%%%%%%%%%%%%%%%%%%%%%%%%%%%%%%%%%%%%%%%%%%%%%%%%%%%%%%%%%%%%%%%%%%%%%%%%%%%%%%%

\end{document}